\documentclass[conf]{new-aiaa}
\usepackage[utf8]{inputenc}

\usepackage{graphicx}
\usepackage{amsmath}
\usepackage[version=4]{mhchem}
\usepackage{siunitx}
\usepackage{longtable,tabularx}
\usepackage[usenames,dvipsnames]{color}
\usepackage{natbib}     % cite author
\usepackage{mathtools}      % for dcases
\usepackage[linesnumbered,ruled,vlined]{algorithm2e}    % algorithm
\usepackage{subcaption}             % subfigure
\usepackage[export]{adjustbox}      % aligning figures
\graphicspath{ {./images/} }
\newcommand{\beq}{\begin{equation}}
\newcommand{\eeq}{\end{equation}}
\newcommand{\overbar}[1]{\mkern 1.5mu\overline{\mkern-1.5mu#1\mkern-1.5mu}\mkern 1.5mu}
\newcommand{\overtilde}[1]{\mkern 1.5mu\widetilde{\mkern-1.5mu#1\mkern-1.5mu}\mkern 1.5mu}

\DeclareMathOperator*{\argmax}{argmax}

\title{Informative Planning of Mobile Sensor Networks in GPS-Denied Environments}

\author{Youngjae Min\footnote{Undergraduate Student, Schoool of Electrical Engineering, yjmin313@kaist.ac.kr}, Soon-Seo Park\footnote{Ph.D. Candidate, Department of Aerospace Engineering, sspark@lics.kaist.ac.kr}, and Han-Lim Choi\footnote{Associate Professor, Department of Aerospace Engineering, hanlimc@kaist.ac.kr}}
\affil{Korea Advanced Institute of Science and Technology, Daejeon, 34141, Republic of Korea}

\begin{document}

\maketitle

\begin{abstract}

This paper considers the problem to plan mobile sensor networks for target localization task in GPS-denied environments. Most researches on mobile sensor networks assume that the states of the sensing agents are precisely known during their missions, which is not feasible under the absence of external infrastructures such as GPS. Thus, we propose a new algorithm to solve this problem by: (i) estimating the states of the sensing agents in addition to the target's through the combination of a particle filter (PF) and extended Kalman filters (EKF) and (ii) involving the uncertainty of the states of the sensing agents in planning the sensor networks based on the combined filters.
This approach does not require any additional internal/external sensors nor the prior knowledge of the surrounding environments.
We demonstrate the limitations of prior works in GPS-denied environments and the improvements from the proposed algorithm through Monte Carlo experiments.

\end{abstract}

%\section{Nomenclature}
%
%{\renewcommand\arraystretch{1.0}
%	% \begin{minipage}{0.48\textwidth}
%	\noindent\begin{longtable*}{@{}l @{\quad=\quad } l@{}}
%		$n_v$   & number of agents (sensors) \\
%		$n_p$   & number of particles in particle filter \\
%		$\theta$& target state \\
%		$\mathrm{x}$     & agent state \\
%		$z$     & measurement of sensor against target \\
%		$u$     & control input \\
%		$h$     & sensor model \\
%		$n$     & sensing noise \\
%		$R$     & covariance matrix for $n$ \\
%		$f$     & motion model of agent\\
%		$\nu$   & movement noise \\
%		$P$     & covariance matrix for $\nu$ \\
%		$dt$    & time step \\
%		$\eta$  & normalization constants \\
%		$\phi$  & angle between the sensor’s boresight and the direction from the sensor to the target \\
%		$g$     & gravitational acceleration \\
%		$I$     & mutual information \\		
%		$H$     & entropy \\
%		$w$     & particle weight \\
%		$\delta$   & dirac delta function \\
%%\overbar{\mu} \overbar{\Sigma} \overtilde{\Sigma} \overtilde{\mu		
%				
%		\multicolumn{2}{l}{\it{super/subscript}} \\
%		$(\cdot)^{(i)}$     & quantity for agent $i$ \\
%		$(\cdot)^t$     & set of quantities from time $1$ to $t$ \\
%		$(\cdot)_t$     & quantity for time $t$ \\
%		$(\cdot)_k$     & quantity for particle $k$
%\end{longtable*}}

\section{Introduction} \label{introduction}

\lettrine{T}echnological advances in mobile sensor networks have enabled the applications of monitoring spatio-temporal phenomena, terrain mapping, and source localization \cite{dunbabin2012robots,choi2010continuous,hoffmann2009mobile}. 
The key technical challenge in the operation of such networked mobile sensor systems is to efficiently deploy and maneuver the mobile sensors to maximize the information about the target systems of interest. 
These kinds of problems can be described as a sensor planning problem that optimizes the utilization of the mobile sensors given current states of the target systems and resources while considering the mobility of the mobile sensors.
% This problem requires an estimation of a distribution over the set of possible states based on noisy measurements acquired by the mobiles sensors.

Information and game theoretic approaches have been extensively explored to solve such problem by efficiently operating mobile sensor networks \cite{hoffmann2009mobile,shen2010game,choi2015potential,lee2018potential}. 
For instance, \citeauthor{hoffmann2009mobile} \cite{hoffmann2009mobile} proposed sensor network planning framework which is scalable and capable of accurately capturing and using information. The mutual information between the sensor measurements and the target state is directly computed using a particle filter representation of the posterior distribution without Gaussian approximation. Mutual information is one popular metric of information gain, which is known as a more fundamental quantity of information than entropy \cite{kolmogorov1956shannon}. 
\citeauthor{choi2015potential} \cite{choi2015potential} and \citeauthor{lee2018potential} \cite{lee2018potential} also adopted mutual information to quantify the amount of the information about the target systems and formulated  the sensor network planning problems into a potential game. The joint strategy fictitious play method is then applied
to obtain a distributed solution that provably converges to a pure
strategy Nash equilibrium.  
\citeauthor{shen2010game} \cite{shen2010game} proposed a negotiable game-theoretic based sensor management method to deal with the requirements of dynamic sensor management and assignment. 

However, such approaches generally assume that they have perfect information and control on the state of each agent in the networks through the internal sensors, such as on-board IMU, and external infrastructures, such as Global Positioning System (GPS) \cite{misra2006global}. The estimation on the position of an agent through GPS is a very simple and powerful method, but GPS could be unavailable in many cases \cite{balamurugan2016survey}. For instance, the GPS signals are normally degraded in complex environments such as mountainous terrains or the places surrounded by tall buildings. Also, the signals are largely attenuated in indoor, underwater, and underground environments. In extreme scenarios, GPS could be down by adversarial attacks, or the sensor networks could be deployed in the space where GPS is not facilitated. To reduce the uncertainty in the states of the sensor networks under those GPS-denied environments, the networks can utilize sensing information on their surrounding environments. UAVs with vision sensors, like cameras and LIDARs, have shown the ability to localize themselves in \cite{hardy2016unmanned,hemann2016long,achtelik2009autonomous}, and \citeauthor{kassas2013observability} \cite{kassas2013observability} have shown the capability for UAVs with low flight attitudes utilizing the signals from cellular communications. However, these approaches focus on single-agent systems, consider a simple mission of arriving at a destination, and/or require prior knowledge on surrounding environments \cite{hardy2016unmanned,hemann2016long,achtelik2009autonomous,he2008planning}.

In this paper, we investigate the target localization problem with mobile sensor networks in GPS-denied environments. There are two key technical challenges: (i) to estimate the states of the sensing agents without any infrastructures (i.e., GPS) and prior knowledge on the environments, (ii) to plan the sensor networks for fast and accurate target localization given the uncertain knowledge on the states of the sensing agents.
We tackle these challenges with a novel approach motivated by the Simultaneous Localization and Mapping (SLAM) problem \cite{montemerlo2002fastslam}. The SLAM problem can be interpreted as a similar target localization problem where a single sensing agent localizes itself and multiple targets given the control inputs. As in the SLAM problem, we integrate two types of Bayesian filters with extension to the multi-agent systems. Those filters are exploited not only for estimating the states of the target and the sensing agents but for evaluating the information gains of future control inputs in the sensor networks planning.
Note that our framework does not require any additional internal/external sensors nor the prior knowledge on the surrounding environments.

The rest of this paper is organized as follows. 
In Sec.~\ref{problem}, we define the target localization problem with mobile sensor networks in GPS-denied environments. The problem is mathematically formulated as an optimization problem maximizing the mutual information between future sensor measurements and the target state. Then, the details of the planning and estimating algorithm are explained in Sec.~\ref{method}. Finally, in Sec.~\ref{simul}, the proposed algorithm is evaluated in comparison to the existing method through Monte Carlo experiments.

\section{Problem Formulation} \label{problem}

\subsection{Basic Settings}

This paper considers a target localization problem. A mobile sensor network, which consists of $n_v$ agents such as UAVs\footnote{In this paper, we use the terms 'agent' and 'UAV' interchangeably.}, tracks a stationary target. At the beginning of the task, only the abstract range of the target's position is given. Each $i^{th}$ agent with its state vector $x^{(i)}_t$ indirectly observes the target's position vector $\theta$ at each time step $t$ by
\beq
\label{eq:observation}
z^{(i)}_t = h^{(i)}(\mathrm{x}^{(i)}_t,\theta) + n^{(i)}_t,
\eeq
where $h^{(i)}$ could be any differentiable function and $n^{(i)}_t \sim \mathcal{N}(0,R^{(i)}_t)$ is the sensing noise following zero-mean Gaussian with covariance matrix $R^{(i)}_t$.
$z^{(i)}_t$ is called a sensor measurement. Unlike the prior works \cite{hoffmann2009mobile,shen2010game,choi2015potential,lee2018potential} which assume perfect controls of sensing agents, we handle the situation in which the controls involve uncertainty. The discrete-time dynamics of the $i^{th}$ agent is
\beq
\label{eq:motion}
\mathrm{x}^{(i)}_t = f^{(i)}(\mathrm{x}^{(i)}_{t-1},u^{(i)}_t) + \nu^{(i)}_t,
\eeq
where $f^{(i)}$ could be any differentiable function, $u^{(i)}_t$ is the control input vector, and $v^{(i)}_t \sim \mathcal{N}(0,P^{(i)}_t)$ is the noise (or uncertainty) in the control following zero-mean Gaussian with covariance matrix $P^{(i)}_t$.
In the rest of the paper, we drop the superscript $(i)$ when denoting the entity concatenated for all agents.

We assume that the initial distributions of the states of the sensing agents are given at the beginning of the mission. This is a common assumption for many problems of system control/planning to begin with such prior knowledge of the system's initial state, and it makes sense since we form the initial topology of the sensor networks before the mission starts.
On the other hand, the information about the target is totally unknown, except the abstract range where it may exist. Thus, the prior distribution of the target's location is set to be uniform over the area. 
%The number of targets is assumed to be one, but it is not a limitation of our algorithm and can be extended to multiple targets.

This work targets two goals in the localization task,
\begin{itemize}
    \item Use successive measurements from the sensor network to estimate the sates of the target and the agents simultaneously.
    \item Plan the motion of each agent at each time step to enable fast and accurate localization.
\end{itemize}
Note that it is important to have accurate estimations on the agents' states as well as the target's. For instance, there may be further missions for the sensor network after finishing the current target localization task. The agents may take additional actions for the localized target or they may return to the original position where they start from. These further missions require accurate knowledge of the agents' states.

\subsection{Mutual Information as Objective Function}

The primary goal of this problem is to localize the target as fast and accurate as possible. Since the frequencies of the sensor measurements and, thus, updates of the estimations are fixed, it is required to minimize the number of total measurements until reaching a sufficient accuracy of the estimations to achieve the fast and accurate localization. Meanwhile, what we can control is only the control inputs for the motion of each agent. Thus, the goal is equivalent to choosing control inputs of the agents at each time step so that they minimize the number of future measurements until having accurate estimations.

While it is very challenging to formulate the number of required future measurements, if not impossible, we can approximate the goal using `mutual information'. In single-step planning, an effective strategy to reduce the number of future measurements is to move the agents to positions in which they take observations with the maximum information on the target. This information can be formulated as the mutual information between the next measurements and the target state. Thus, the problem is interpreted as solving the following optimization problem:
\beq
\label{eq:optimization}
u^*_{t+1} = \argmax_{u_{t+1}} I(z_{t+1}; \theta \mid z^t, u^{t+1}, \mathrm{x}_1),
\eeq
where $I(A;B|C)$ denotes the mutual information between $A$ and $B$ given $C$. The superscript $t$ means the entity concatenated from time 1 to $t$.

\section{Methodology} \label{method}

\begin{figure}[t!]
    \centering
    \includegraphics[width=0.9\textwidth]{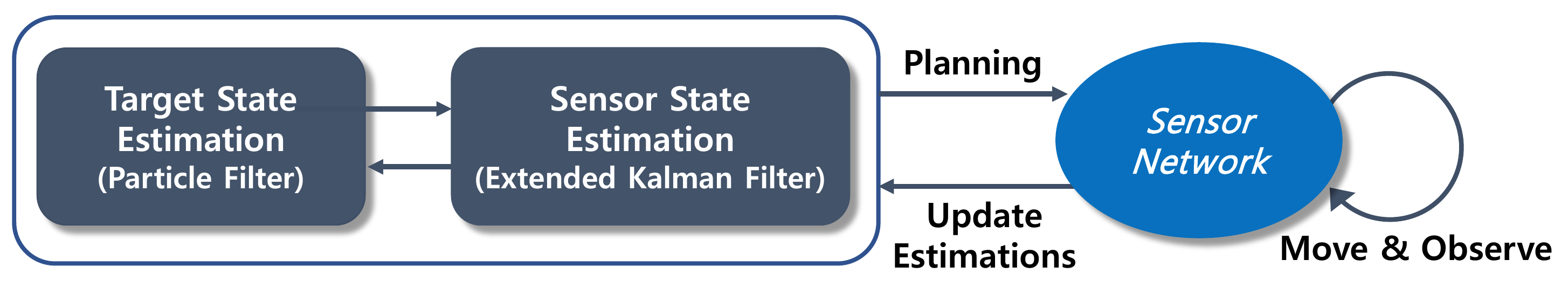}
    \caption{System overview}
    \label{fig:overview}
\end{figure}

In this section, we explore how the overall system operates in detail. The overview of the whole system is shown in Fig.~\ref{fig:overview}.
The problem defined in the previous section requires the state estimations of the sensing agents as well as the target. The estimations are performed based on the prior knowledge of the sensing system with a finite set of measurements obtained from the sensing agents, which naturally fits into Bayesian state estimation (i.e., Bayesian filtering \cite{chen2003bayesian}) formulation. Bayesian filtering formulation provides us a way to evaluate the mutual information to be maximized in \eqref{eq:optimization} at each time step. We combine a particle filter and extended Kalman filters to estimate the states of the target and the sensing agents simultaneously.
While the target state estimation and the sensor state estimation are done with different types of filters, their updates depend on each other. Based on the estimations, the sensor network plans the next control inputs which maximize the objective function defined in \eqref{eq:optimization}. Then, the network updates the agents' states with the control input according to \eqref{eq:motion}. New observations are made with the new states following \eqref{eq:observation} and the estimations are updated reflecting the new observations.

\subsection{Bayesian Filtering}
The overall system is assumed to satisfy the Markov property. Hence, recursive updates through Bayes' rule are used to incorporate the sensor measurements and the control inputs in the state estimations along successive times. 
For the simultaneous estimation on both the target's and the agents' sates, we utilize the factorization of their joint posterior,
\beq
p(\theta, \mathrm{x}_t\mid z^t, u^t,\mathrm{x}_1) = p(\theta\mid z^t, u^t,\mathrm{x}_1) \prod^{n_v}_{i=1} p(\mathrm{x}^{(i)}_t\mid \theta, z^t, u^t,\mathrm{x}_1),
\eeq
as in \cite{murphy2000bayesian,montemerlo2002fastslam}. Instead of estimating the joint distribution directly, this factorization is utilized through constructing a filter for $\theta$ and, then, filters for $\mathrm{x}_t$ depending on the filter for $\theta$.

To form the dependency, first, a particle filter is employed to estimate the target state. The particle filter is a non-parametric Monte Carlo method for Bayesian estimation, which enables nonlinear updates for non-Gaussian probability distributions of target estimations. Although the particle filter typically incurs more computational cost than other estimation methods based on Gaussian approximation, it generally results in more accurate estimations when enough particles are employed. The $n_p$ particles represent the target state in the form of a probability mass function as:
%To form the dependency, first, a particle filter can be employed to estimate the target state by\\
\beq
\label{eq:particlefilter}
p(\theta\mid z^t, u^t,\mathrm{x}_1) \approx \sum^{n_p}_{k=1} w_{t,k}\delta(\theta-\theta_{t,k}) \quad\text{s.t.}\quad \sum^{n_p}_{k=1} w_{t,k} = 1,
\eeq
where $w_{t,k}$ and $\theta_{t,k}$ are the weight and the state for the $k^{th}$ particle, respectively. $\delta$ denotes the Dirac delta function.

Then, instead of considering the whole infinite $\theta$-space, we just need to deal with the finite number of filters for $\mathrm{x}$ conditioned on each particle. Here, the extended Kalman filter (EKF) is used for computational efficiency. The EKF represents $\mathrm{x}$ as:
\beq
p(\mathrm{x}^{(i)}_t\mid \theta_{t,k}, z^t, u^t,\mathrm{x}_1)
\sim \mathcal{N}(\mathrm{x}^{(i)}_t; \mu^{(i)}_{t,k},\Sigma^{(i)}_{t,k}),
\eeq
where $N(A;B,C)$ denotes the multivariate Gaussian probability density function of $A$ with the mean vector $B$ and the covariance matrix $C$.
%$\mu^{(i)}_{t,k}$ and $\Sigma^{(i)}_{t,k}$ are the mean vector and covariance matrix of the Gaussian distribution for the $i^{th}$ agent's state conditioned on the $k^{th}$ particle.
Note that since the exact locations of the sensing agents are given in the beginning, the initial covariances of their state estimations are zero, and the linear approximation steps in the EKF are accurate.

\subsection{Sensor State Estimation} \label{sensor_state_estimation}

Sensor state estimations are recursively updated through the EKF.
It starts from $p(\mathrm{x}^{(i)}_t\mid \theta_{t,k}, z^t, u^t,\mathrm{x}_1)$ to induce $p(\mathrm{x}^{(i)}_{t+1}\mid \theta_{t,k}, z^{t+1}, u^{t+1},\mathrm{x}_1)$ based on the two-stage belief updates. In the first stage, the belief is updated prior to reflecting the new sensor measurements. We compute the likelihood of the $i^{th}$ sensor's state with the new control inputs, $u_{t+1}$, as below:
\beq
\label{eq:EKFprior}
\begin{split}
p(\mathrm{x}^{(i)}_{t+1}\mid \theta_{t,k}, z^t, u^{t+1},\mathrm{x}_1) & =
\int
\underbrace{p(\mathrm{x}^{(i)}_{t+1}\mid \mathrm{x}^{(i)}_t,u^{(i)}_{t+1})} _{\sim\mathcal{N}(\mathrm{x}^{(i)}_{t+1};f^{(i)}(\mathrm{x}^{(i)}_t,u^{(i)}_{t+1}),P^{(i)}_{t+1})} \underbrace{p(\mathrm{x}^{(i)}_t\mid \theta_{t,k}, z^t, u^t,\mathrm{x}_1)}
_{\sim\mathcal{N}(\mathrm{x}^{(i)}_t; \mu^{(i)}_{t,k},\Sigma^{(i)}_{t,k})}
d \mathrm{x}^{(i)}_t
\approx \mathcal{N}(\mathrm{x}^{(i)}_{t+1};\overbar{\mu}^{(i)}_{t+1,k},\overbar{\Sigma}^{(i)}_{t+1,k})
\end{split}
\eeq
with
\begin{equation*}
\overbar{\mu}^{(i)}_{t+1,k} = f^{(i)}(\mu^{(i)}_{t,k},u^{(i)}_{t+1})\;,\;\;\;
\overbar{\Sigma}^{(i)}_{t+1,k} = F_\mathrm{x}\Sigma^{(i)}_{t,k}F^{T}_\mathrm{x}+P^{(i)}_{t+1}\;,\quad\text{where}\quad
F_\mathrm{x} = \nabla_\mathrm{x} f^{(i)}(\mu^{(i)}_{t,k},u^{(i)}_{t+1})
\end{equation*}
Here, the first equality holds from Bayes' rule with the Markov property, and the stochastic motion model in \eqref{eq:motion} is applied to compute the integral.

In the second stage, the posterior state estimation is developed from the prior estimation by reflecting the new sensor measurements, $z_{t+1}$. It corrects the prior state belief in \eqref{eq:EKFprior} with the new measurements as
\beq
\label{eq:EKFposterior}
\begin{split}
p(\mathrm{x}^{(i)}_{t+1}\mid \theta_{t,k}, z^{t+1}, u^{t+1},\mathrm{x}_1) &= \eta^{(i)}_{\mathrm{x},t+1}
\underbrace{p(z^{(i)}_{t+1}\mid \mathrm{x}^{(i)}_{t+1},\theta_{t,k})} _{\sim\mathcal{N}(z^{(i)}_{t+1};h^{(i)}(\mathrm{x}^{(i)}_{t+1},\theta_{t,k}),R^{(i)}_{t+1})}
\underbrace{p(\mathrm{x}^{(i)}_{t+1}\mid \theta_{t,k}, z^t, u^{t+1},\mathrm{x}_1)}
_{\sim\mathcal{N}(\mathrm{x}^{(i)}_{t+1};\overbar{\mu}^{(i)}_{t+1,k},\overbar{\Sigma}^{(i)}_{t+1,k})}
\approx \mathcal{N}(\mathrm{x}^{(i)}_{t+1};\mu^{(i)}_{t+1,k},\Sigma^{(i)}_{t+1,k})
\end{split}
\eeq
with
\begin{equation*}
%\label{eq:EKFposterior_detail}
\mu^{(i)}_{t+1,k} = \overbar{\mu}^{(i)}_{t+1,k} + K(z^{(i)}_{t+1}-h^{(i)}(\overbar{\mu}^{(i)}_{t+1,k},\theta_{t,k}))\;,\;\;\;
\Sigma^{(i)}_{t+1,k} = (I-K H_\mathrm{x})\overbar{\Sigma}^{(i)}_{t+1,k},
\end{equation*}
where
\begin{equation*}
%\mu^{(i)}_{t+1,k} & = \overbar{\mu}^{(i)}_{t+1,k} + K(z^{(i)}_{t+1}-h^{(i)}(\overbar{\mu}^{(i)}_{t+1,k},\theta_{t,k}))\\
%\Sigma^{(i)}_{t+1,k} & = (I-KH_\mathrm{x})\overbar{\Sigma}^{(i)}_{t+1,k}\\
K = \overbar{\Sigma}^{(i)}_{t+1,k}H^{T}_\mathrm{x}(H_\mathrm{x}\overbar{\Sigma}^{(i)}_{t+1,k}H^{T}_\mathrm{x}+R^{(i)}_{t+1})^{-1}\;,\;\;\;
H_\mathrm{x} = \nabla_\mathrm{x} h^{(i)}(\overbar{\mu}^{(i)}_{t+1,k},\theta_{t,k}).
\end{equation*}
$\eta^{(i)}_{\mathrm{x},t+1}$ is a normalizing constant.
Again, the computation is based on the Bayes' rule and the Markov property, and the sensor measurement model in \eqref{eq:observation} is applied.
%\eqref{eq:EKFprior} and \eqref{eq:EKFposterior} define the belief dynamics.
Note that the second term, called the innovation term, in the representation of $\mu^{(i)}_{t+1,k}$ depends on the measurement $z_{t+1}^{(i)}$ with the Kalman gain $K$.
%$\Sigma^{(i)}_{t+1,k}$ evolves given the current covariance of the sensing agent, regardless of the measurement. This represents the uncertainty of the agent states.

\subsection{Target State Estimation} \label{target_state_estimation}

We employ a particle filter to recursively update the target state estimation. As the filter consists of finite particles and corresponding weights, we need two types of updates, one for the particles and the other for the weights. The former is simple.
Since the filter estimates the stationary target, the particles are motionless as well. This implies $\theta_{t+1,k} = \theta_{t,k}$, unless resampling occurs. Resampling increases the accuracy of the discrete approximation in \eqref{eq:particlefilter} by rearranging particles to more probable regions. In each time step, when the effective number of particles $N_{eff}:= 1/(\sum w_{t+1,k}^2) < n_p/2$, the low variance resampling method \cite{thrun2005probabilistic} is applied.

For the update of the weights, Bayes' rule is applied as:
\beq
p(\theta\mid z^{t+1}, u^{t+1},\mathrm{x}_1) \propto p(z_{t+1}\mid \theta, z^t, u^{t+1},\mathrm{x}_1) p(\theta\mid z^t, u^t,\mathrm{x}_1)
\approx \sum^{n_p}_{k=1} p(z_{t+1}\mid \theta_{t,k}, z^t, u^{t+1},\mathrm{x}_1) w_{t,k}\delta(\theta-\theta_{t,k}).
\eeq
The last approximation comes from \eqref{eq:particlefilter} with the characteristics of the unit impulse of the Dirac delta function. Then, the new weights are obtained by rescaling the original weights and normalizing them. The relative scaling factor of each weight is calculated as
\beq
	p(z_{t+1}\mid \theta_{t,k}, z^t, u^{t+1},\mathrm{x}_1)  = \prod^{n_v}_{i=1}
	\int
	\underbrace{p(z^{(i)}_{t+1}\mid \mathrm{x}^{(i)}_{t+1},\theta_{t,k})} _{\sim\mathcal{N}(z^{(i)}_{t+1};h^{(i)}(\mathrm{x}^{(i)}_{t+1},\theta_{t,k}),R^{(i)}_{t+1})} \underbrace{p(\mathrm{x}^{(i)}_{t+1}\mid \theta_{t,k}, z^t, u^{t+1},\mathrm{x}_1)}
	_{\sim\mathcal{N}(\mathrm{x}^{(i)}_{t+1};\overbar{\mu}^{(i)}_{t+1,k},\overbar{\Sigma}^{(i)}_{t+1,k})}
	d \mathrm{x}^{(i)}_{t+1}
	 = \prod^{n_v}_{i=1} \mathcal{N}(z^{(i)}_{t+1};\overtilde{\mu}^{(i)}_{t+1,k},\overtilde{\Sigma}^{(i)}_{t+1,k})
\eeq
with
\begin{equation*}
\overtilde{\mu}^{(i)}_{t+1,k} = h^{(i)}(\overbar{\mu}^{(i)}_{t+1,k},\theta_{t,k})\;,\;\;\;
\overtilde{\Sigma}^{(i)}_{t+1,k} = H_\mathrm{x}\overbar{\Sigma}^{(i)}_{t+1,k}H^{T}_\mathrm{x}+R^{(i)}_{t+1}\;,\quad\text{where}\quad
H_\mathrm{x} = \nabla_\mathrm{x} h^{(i)}(\overbar{\mu}^{(i)}_{t+1,k},\theta_{t,k}).
\end{equation*}
Note that $H_\mathrm{x}$ takes different value for each $(i,k,t)$ combination. Finally, the particle filter is updated as
\beq
\label{eq:pf_update}
	w_{t+1,k} = \eta_{w,t+1} \Big[\prod^{n_v}_{i=1} \mathcal{N}(z^{(i)}_{t+1};\overtilde{\mu}^{(i)}_{t+1,k},\overtilde{\Sigma}^{(i)}_{t+1,k})\Big] w_{t,k}\;,\;\;\;
	\theta_{t+1,k} = \theta_{t,k},
\eeq
where $\eta_{w,t+1}$ is a normalizing constant to let the total sum of the weights be 1. When the resampling happens, the particles are rearranged and the weights are set uniformly.

\subsection{Evaluation of Mutual Information}

The mutual information in (\ref{eq:optimization}) is defined as
\beq
\label{eq:mutual_information}
I(z_{t+1};\theta\mid z^t,u^{t+1},\mathrm{x}_1) := H(z_{t+1}) - H(z_{t+1}\mid\theta)\quad\text{given}\quad z^t,u^{t+1},\mathrm{x}_1,
\eeq
where $H(A)$ denotes the entropy of $A$, and $H(A\mid B)$ denotes the conditional entropy of $A$ conditioned on $B$.
The first term in RHS is, by definition,
\beq
\label{eq:H(z)}
H(z_{t+1}) = - \int p(z_{t+1}\mid z^t,u^{t+1},\mathrm{x}_1)\log p(z_{t+1}\mid z^t,u^{t+1},\mathrm{x}_1) dz_{t+1},
\eeq
\vspace{-0.4cm}\\
where
\vspace{-0.4cm}
\beq
\label{eq:p(z)}
\begin{split}
p(z_{t+1}\mid z^t,u^{t+1},\mathrm{x}_1)
& = \int p(z_{t+1}\mid \theta,z^t,u^{t+1},\mathrm{x}_1) p(\theta\mid z^t,u^t,\mathrm{x}_1)d\theta\\
& \approx \sum^{n_p}_{k=1} w_{t,k} p(z_{t+1}\mid \theta_{t,k},z^t,u^{t+1},\mathrm{x}_1)\\
% & = \iint p(z_{t+1}\mid \mathrm{x}_{t+1},\theta) p(\mathrm{x}_{t+1}\mid \theta, z^t, u^{t+1},\mathrm{x}_1) p(\theta\mid z^t,u^t,\mathrm{x}_1)d\theta d \mathrm{x}_{t+1}\\
% & \approx \sum^{n_p}_{k=1} w_{t,k} \prod^{n_v}_{i=1}
% \int
% \underbrace{p(z^{(i)}_{t+1}\mid \mathrm{x}^{(i)}_{t+1},\theta_{t,k})} _{\sim\mathcal{N}(z^{(i)}_{t+1};g(\mathrm{x}^{(i)}_{t+1},\theta_{t,k}),R_{t+1})} \underbrace{p(\mathrm{x}^{(i)}_{t+1}\mid \theta_{t,k}, z^t, u^{t+1},\mathrm{x}_1)}
% _{\sim\mathcal{N}(\mathrm{x}^{(i)}_{t+1};\overbar{\mu}^{(i)}_{t+1,k},\overbar{\Sigma}^{(i)}_{t+1,k})}
% d \mathrm{x}^{(i)}_{t+1}\\
& \approx \sum^{n_p}_{k=1} w_{t,k} \prod^{n_v}_{i=1} \mathcal{N}(z^{(i)}_{t+1};\overtilde{\mu}^{(i)}_{t+1,k},\overtilde{\Sigma}^{(i)}_{t+1,k}).
\end{split}
\eeq
The evaluation of \eqref{eq:H(z)} with \eqref{eq:p(z)} is computationally very heavy and not scalable since the integration in \eqref{eq:H(z)} is over $\mathbb{R}^{n_v}$. Thus, for computational efficiency, we use Gaussian approximation technique. \eqref{eq:p(z)} is approximated as a Gaussian distribution with the same mean and covariance of itself. Then, its entropy can be computed analytically. The mean, $\hat{\mu}$, and the covariance, $\hat{\Sigma}$, of (\ref{eq:p(z)}) are
\beq
\label{eq:gaussian_approx}
	\hat{\mu} = \sum^{n_p}_{k=1} w_{t,k}
	\begin{bmatrix}
		\overtilde{\mu}^{(1)}_{t+1,k} & \cdot\cdot\cdot & \overtilde{\mu}^{(n_v)}_{t+1,k} \end{bmatrix}^T\;,\;\;\;
	\hat{\Sigma} : \hat{\Sigma}_{(i,j)} =
	\begin{dcases}
		\sum^{n_p}_{k=1} w_{t,k} (\overtilde{\mu}^{(i)}_{t+1,k}-\hat{\mu}^{(i)}) (\overtilde{\mu}^{(j)}_{t+1,k}-\hat{\mu}^{(j)}) & \text{if}\,i\neq j\\
		\sum^{n_p}_{k=1} w_{t,k} (\overtilde{\Sigma}^{(i)}_{t+1,k} + (\overtilde{\mu}^{(i)}_{t+1,k})^2) - (\hat{\mu}^{(i)})^2 & \text{if}\,i=j
	\end{dcases}.
\eeq
Then, \eqref{eq:H(z)} is approximated to the analytic evaluation as
\beq
\label{eq:H(z)_eval}
H(z_{t+1}) \approx \frac{1+\log(2\pi)+\log|\hat{\Sigma}|}{2}.
\eeq

The second term in the RHS of (\ref{eq:mutual_information}) is
\vspace{-0.2cm}
\beq
\label{eq:H(z|theta)}
\begin{split}
H(z_{t+1}\mid\theta) & = - \iint p(z_{t+1}\mid\theta,z^t,u^{t+1},\mathrm{x}_1)\log p(z_{t+1},\theta\mid z^t,u^{t+1},\mathrm{x}_1) d\theta dz_{t+1}\\
& \approx -\sum^{n_p}_{k=1} w_{t,k} \sum^{n_v}_{i=1}
\int p(z^{(i)}_{t+1}\mid\theta_{t,k},z^t,u^{t+1},\mathrm{x}_1)\log p(z^{(i)}_{t+1}\mid\theta_{t,k},z^t,u^{t+1},\mathrm{x}_1) dz^{(i)}_{t+1}\\
& \approx \sum^{n_p}_{k=1} w_{t,k} \sum^{n_v}_{i=1} \frac{1+\log(2\pi)+\log|\overtilde{\Sigma}^{(i)}_{t+1,k}|}{2},
\end{split}
\eeq
\vspace{-0.5cm}\\
because
\vspace{-0.2cm}
\beq
\begin{split}
p(z^{(i)}_{t+1}\mid\theta_{t,k},z^t,u^{t+1},\mathrm{x}_1)
& = \int p(z^{(i)}_{t+1}\mid \mathrm{x}^{(i)}_{t+1},\theta_{t,k}) p(\mathrm{x}^{(i)}_{t+1}\mid\theta_{t,k},z^t,u^{t+1},\mathrm{x}_1) d \mathrm{x}^{(i)}_{t+1}
\approx \mathcal{N}(z^{(i)}_{t+1};\overtilde{\mu}^{(i)}_{t+1,k},\overtilde{\Sigma}^{(i)}_{t+1,k}).
\end{split}
\eeq
Substituting (\ref{eq:H(z)_eval}) and (\ref{eq:H(z|theta)}) into (\ref{eq:mutual_information}) and disregarding the constant terms, (\ref{eq:optimization}) is reduced to
\beq
\label{eq:optimization_approx}
u^*_{t+1} \approx \argmax_{u_{t+1}} \bigg( \log|\hat{\Sigma}| - \sum^{n_p}_{k=1} w_{t,k} \sum^{n_v}_{i=1} \log|\overtilde{\Sigma}^{(i)}_{t+1,k}|\bigg).
\eeq
Then, the optimization in (\ref{eq:optimization_approx}) is solved numerically.

\begin{algorithm}[H]
	\SetAlgoLined
	\KwResult{$\mu_{t+1},\Sigma_{t+1},w_{t+1},\theta_{t+1},u^*_{t+1}$}
	Given $\mu_t,\Sigma_t,w_t,\theta_t$\;
	\tcp{Plan next step}
	$I_{max}=0$\;
	\For{$u_{t+1}$ in action set}{
		\For{$i=1:n_v$}{
			\For{$k=1:n_p$}{
				update $\overbar{\mu}^{(i)}_{t+1,k}, \overbar{\Sigma}^{(i)}_{t+1,k}, \overtilde{\mu}^{(i)}_{t+1,k}, \overtilde{\Sigma}^{(i)}_{t+1,k}$ (see Sec.~\ref{sensor_state_estimation}, Sec.~\ref{target_state_estimation})\;
			}
		}
		update $\hat{\mu}, \hat{\Sigma}$ using (\ref{eq:gaussian_approx})\;
		$I = \log|\hat{\Sigma}| - \sum^{n_p}_{k=1} w_{t,k} \sum^{n_v}_{i=1} \log|\overtilde{\Sigma}^{(i)}_{t+1,k}|$\;
		\If{$I>I_{max}$}{
			$I_{max}=I$\;
			$u^*_{t+1}=u_{t+1}$\;
			save $\overbar{\mu}_{t+1,k}, \overbar{\Sigma}_{t+1,k}, \overtilde{\mu}_{t+1,k}, \overtilde{\Sigma}_{t+1,k}$ \;
		}
	}
	\tcp{Move $\&$ observe}
	Move to $x_{t+1}$ with $u^*_{t+1}$ using (\ref{eq:motion})\;
	Obtain $z_{t+1}$ at $x_{t+1}$ using (\ref{eq:observation})\;
	\tcp{Update filters}
	\For{$k=1:n_p$}{
		\For{$i=1:n_v$}{
			update $\mu^{(i)}_{t+1,k}\Sigma^{(i)}_{t+1,k}$ (see Sec.~\ref{sensor_state_estimation})\;
		}
		update $w_{t+1,k},\theta_{t+1,k}$ using (\ref{eq:pf_update})\;
	}
	\caption{Planning with uncertain motion model}
	\label{alg:single_step}
\end{algorithm}

\subsection{Summary of Algorithm}

Algorithm~\ref{alg:single_step} summarizes the single step of the system described above. 
Our system consists of three parts: planning (line 1-16), execution (line 17-18), and estimation (line 19-24). 
In the planning part, we optimize the one-step-ahead control inputs of the sensor network using the information of the target, represented by particles, and the agents, represented by Gaussian distributions, given from the previous step. 
Every control input that each agent can take is compared by evaluating the information gain for the target to be obtained when the corresponding control input is executed.
%control input evaluated based on alternating minization.
In the execution part, optimized control inputs are executed and new sensor measurements are obtained at the new states.
In the last part, the Bayesian filters for the target and the agents are updated using the measurements.
This process is repeated until the mission ends.

% \subsection{Alternating Optimization}

% The objective function in (\ref{eq:optimization_approx}) contains many computations such as (\ref{eq:gaussian_approx}) and those in Sec.~\ref{sensor_state_estimation} and Sec.~\ref{target_state_estimation}. 

\section{Simulation Results} \label{simul}

In this section, the proposed method is evaluated through Monte Carlo experiments. In addition to our proposed algorithm, the informative planning algorithm in \cite{hoffmann2009mobile} is implemented to demonstrate the limitations of disregarding the uncertainty in the motion model and evaluate the improvements made by the proposed algorithm. The algorithm in the prior work uses only a particle filter for target state estimation and believes that the motion model is accurate. We denote the algorithm as 'PF-only' algorithm. It is equivalent to the proposed algorithm when the motion noise is set to zero.

For simulation, we choose particular sensor and motion models. Note that any nonlinear differentiable models can be incorporated since the constructed algorithm uses the general abstractions, $h^{(i)}$ and  $f^{(i)}$ for the sensor and motion models. Also, we use the same sensor and motion models for every agent for simplicity. when the agents are different types of UAVs employing different types of sensors, we can easily adopt different models for each sensor.

\begin{figure}[t]
	\centering
	\begin{subfigure}[b]{0.45\textwidth}
		\centering
		\includegraphics[width=\textwidth]{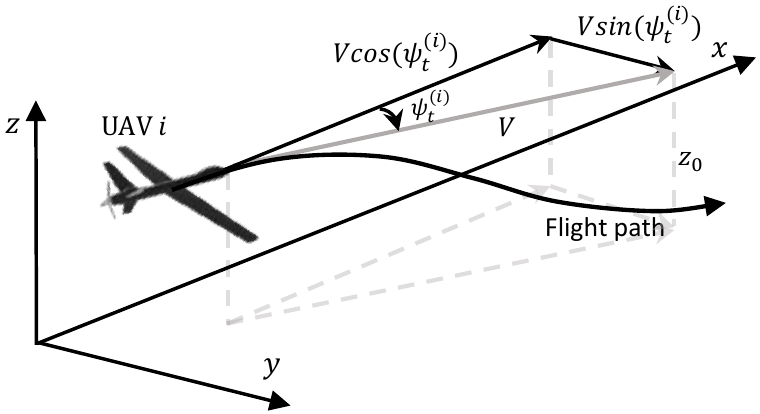}
		\caption{}
		\label{fig:motion}
	\end{subfigure}
	\hspace{1cm}
	\begin{subfigure}[b]{0.45\textwidth}
		\centering
		\includegraphics[width=\textwidth]{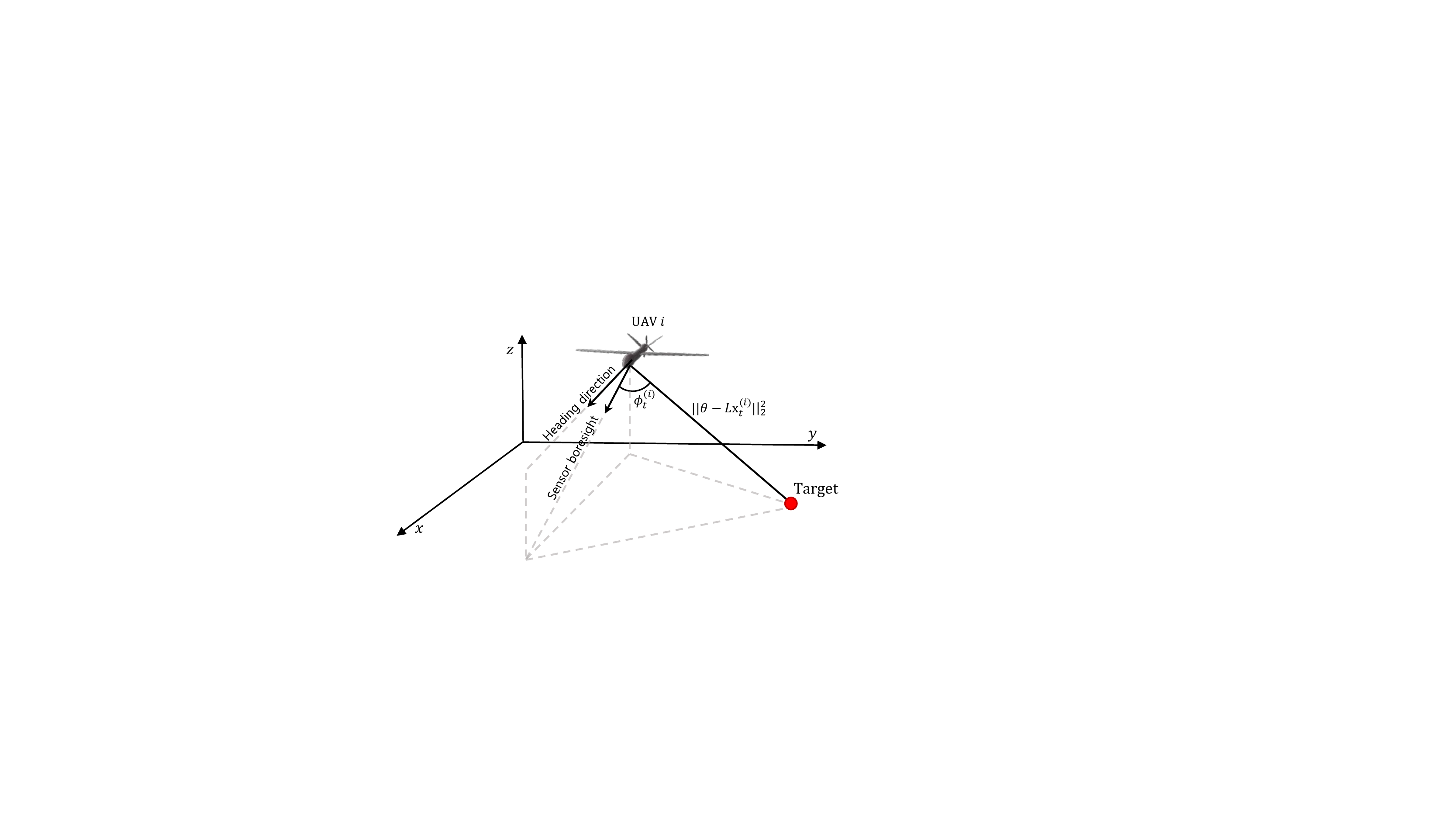}
		\caption{}
		\label{fig:sensor}
	\end{subfigure}
	\caption{Graphical representation of (a) the fixed-wing UAV kinematic model, and (b) sensing geometry for target tracking}
	\label{fig:model}
\end{figure}

\subsection{Sensor Model}

We assume that each agent $i$ measures signal-to-noise ratio (SNR) \cite{skolnik1970radar} of the signal from the target, as shown in Fig.~\ref{fig:model}(b), by

\beq
\label{eq:sensor_model}
h^{(i)}(\mathrm{x}^{(i)}_t,\theta) = \dfrac{\alpha \gamma^{-(\phi^{(i)}_t)^2}} {||\theta-L\mathrm{x}^{(i)}_t||^2_2+\beta},
\eeq
where $\alpha, \beta,$ and $\gamma$ are set to model the SNR of the sensor. ($\alpha=1000, \beta=100, \gamma=3.375$ in this simulation) This model includes both the bearing, $\phi^{(i)}_t$, and the range, $||\theta-L\mathrm{x}^{(i)}_t||$, terms. Note that when $\gamma$ is set to 1, (\ref{eq:sensor_model}) is reduced to the widely used quasi-range measurement model. \cite{williams2007approximate,lee2014efficient} The variance of the Gaussian random noise, $R^{(i)}_t$, is set to 2 for every $i$ and $t$.

\subsection{Motion Model}

We use the dynamics of fixed-wing UAV as in \cite{owen2015implementing} while fixing the altitude at which a UAV is flying and the speed, $V=$1m/s. In this model, there are three state variables, the $(x, y)$ coordinates and the heading angle of a UAV as
\beq
\mathrm{x}^{(i)}_t=
\begin{pmatrix}
x^{(i)}_t & y^{(i)}_t & \psi^{(i)}_t
\end{pmatrix}^T.
\eeq
With this state vector, the motion model is represented as:
\beq
f^{(i)}(\mathrm{x}^{(i)}_t,u^{(i)}_{t+1}) = \mathrm{x}^{(i)}_t +
\begin{pmatrix}
V \cos(\psi^{(i)}_t) &
V \sin(\psi^{(i)}_t) &
\dfrac{\text{g}}{V} \tan(u^{(i)}_{t+1})
\end{pmatrix}^T dt.
\eeq
The control input $u$ (i.e., bank angle of the UAV) is bounded under the condition
\begin{equation}
{\lvert}u{\rvert} \leq u_{max}.
\end{equation}
Assuming a coordinated turn, and given the boundedness of the bank angle $u$, the minimum turn radius $r_{min}$ that the UAV can fly is given by
\begin{equation}
r_{min} = \dfrac{V^2}{g\tan(u_{max})}.
\end{equation}
In the simulation, we set the $u_{max}$ to satisfy $r_{min}=3m$.

\subsection{Performance}

Monte Carlo experiments are performed by 100 trials for each of 6 different noise levels in the motion model. With the reference covariance matrix $P_0 = diag(\sigma^2_x, \sigma_y^2, \sigma_\psi^2)$, where $\sigma_x=$ 0.05 m, $\sigma_y=$ 0.05 m, and $\sigma_\psi=$ 0.0436 rad ($=2.5^\circ$), the 6 noise levels are set with the corresponding covariance matrices $0, 0.5P_0, P_0, 2P_0, 4P_0,$ and $6P_0$. In each experiment, we fix the topology of the sensor network that consists of four agents and limit the searching space as a 40m$\times$40m region. At each time step of each trial, the proposed algorithm and PF-only method are given with the same noises in sensor measurements and movements of the agents for a fair comparison.

\begin{figure}[t]
	\centering
	\begin{subfigure}[b]{0.33\textwidth}
		\centering
		\includegraphics[width=\textwidth]{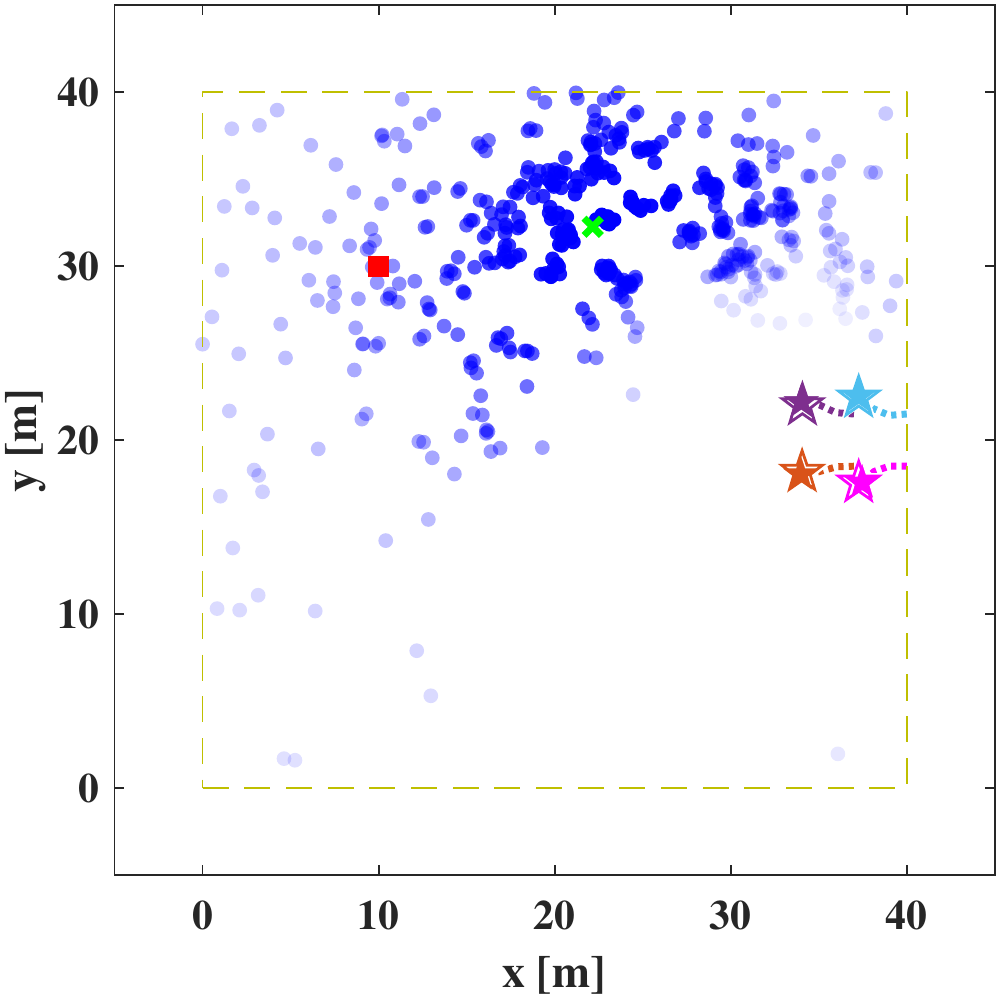}
		\caption{$t=3s$}
		\label{fig:2P_h3}
	\end{subfigure}
	\begin{subfigure}[b]{0.33\textwidth}
		\centering
		\includegraphics[width=\textwidth]{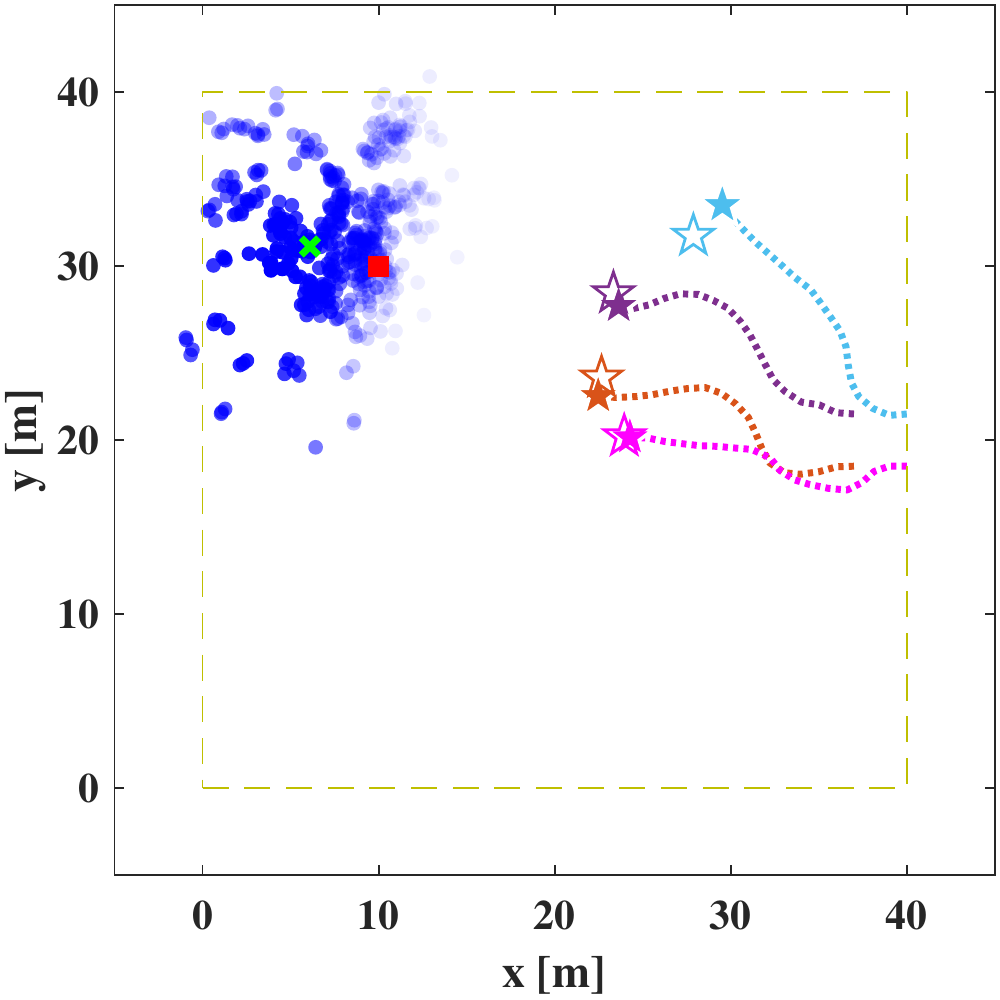}
		\caption{$t=17s$}
		\label{fig:2P_h17}
	\end{subfigure}
	\begin{subfigure}[b]{0.33\textwidth}
		\centering
		\includegraphics[width=\textwidth]{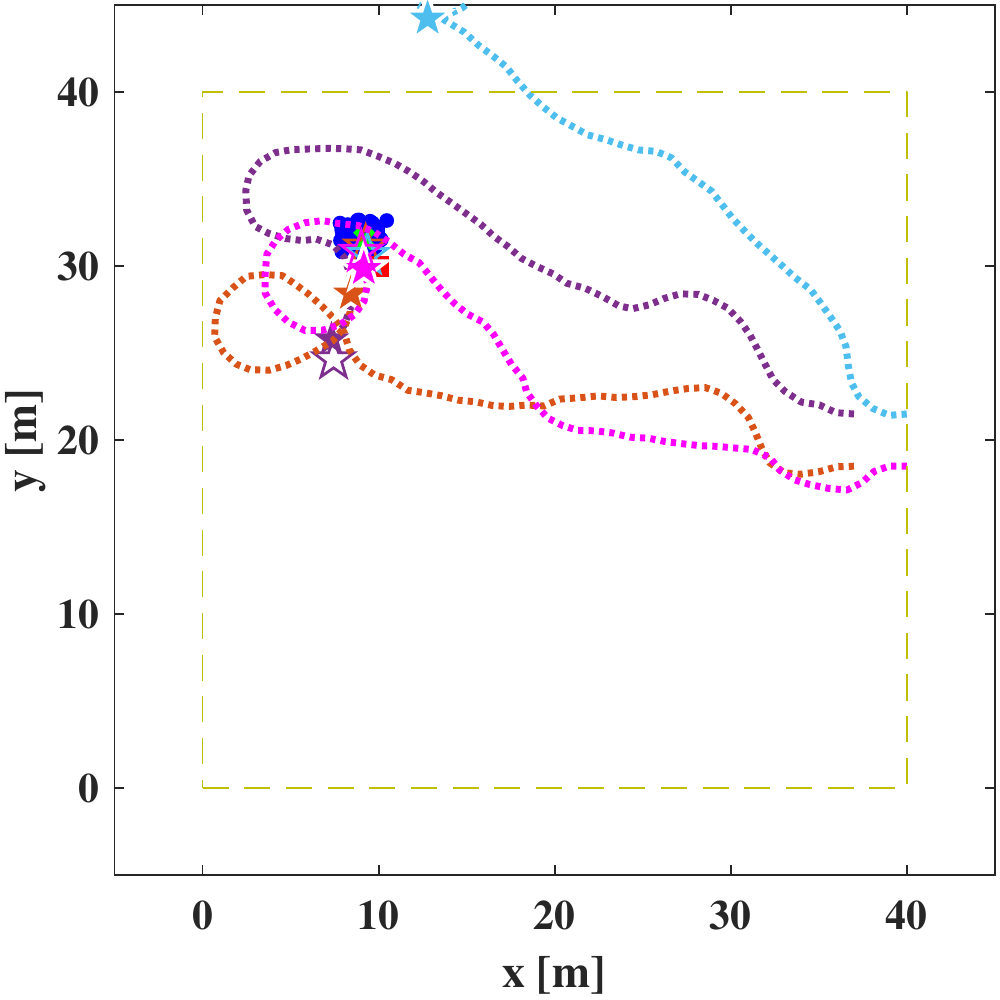}
		\caption{$t=55s$}
		\label{fig:2P_h55}
	\end{subfigure}
	\begin{subfigure}[b]{0.33\textwidth}
		\centering
		\includegraphics[width=\textwidth]{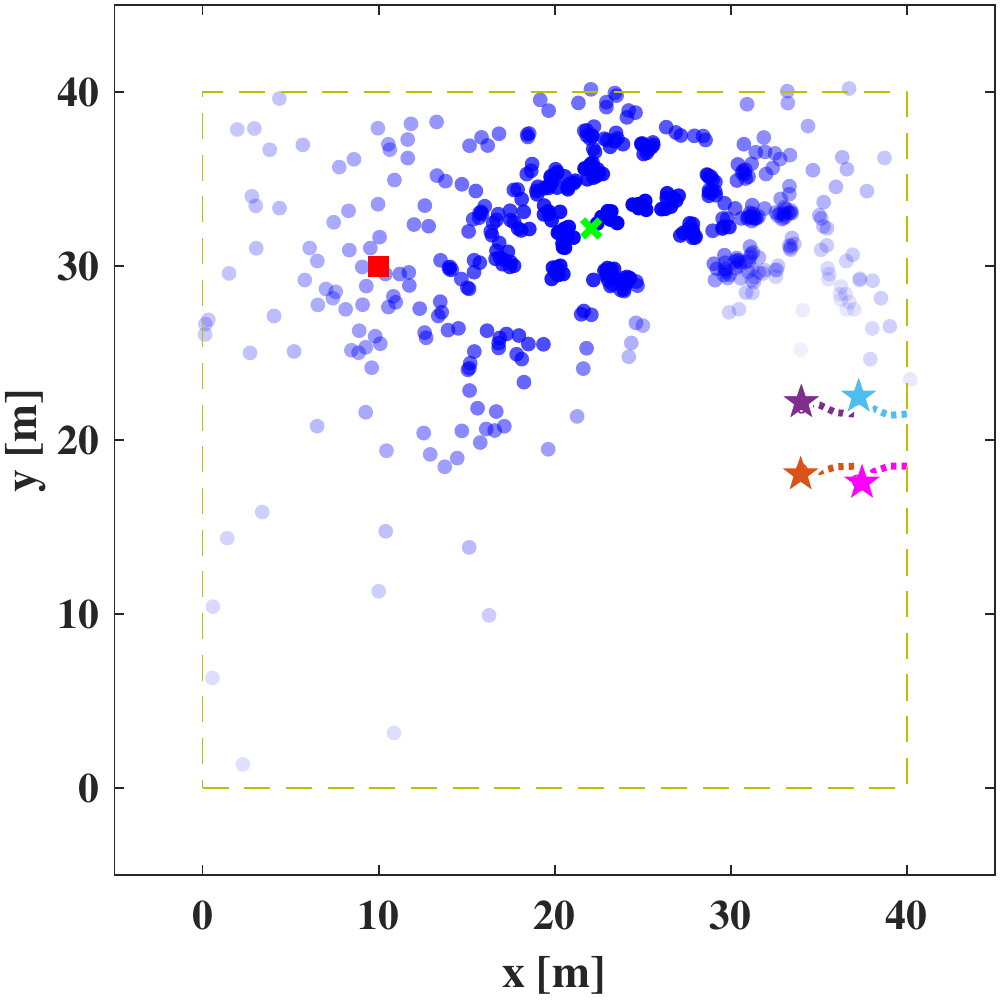}
		\caption{$t=3s$}
		\label{fig:y 2P_o3}
	\end{subfigure}
	\begin{subfigure}[b]{0.33\textwidth}
		\centering
		\includegraphics[width=\textwidth]{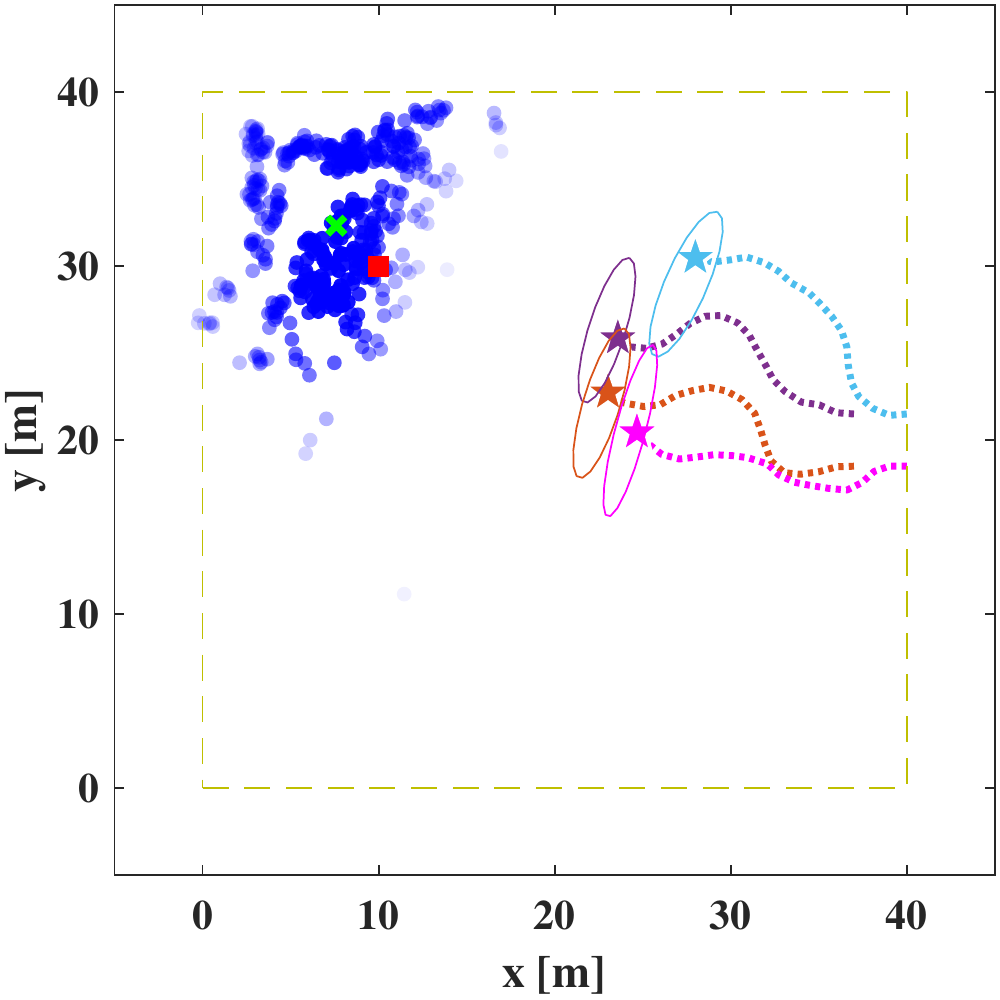}
		\caption{$t=17s$}
		\label{fig:2P_o17}
	\end{subfigure}
	\begin{subfigure}[b]{0.33\textwidth}
		\centering
		\includegraphics[width=\textwidth]{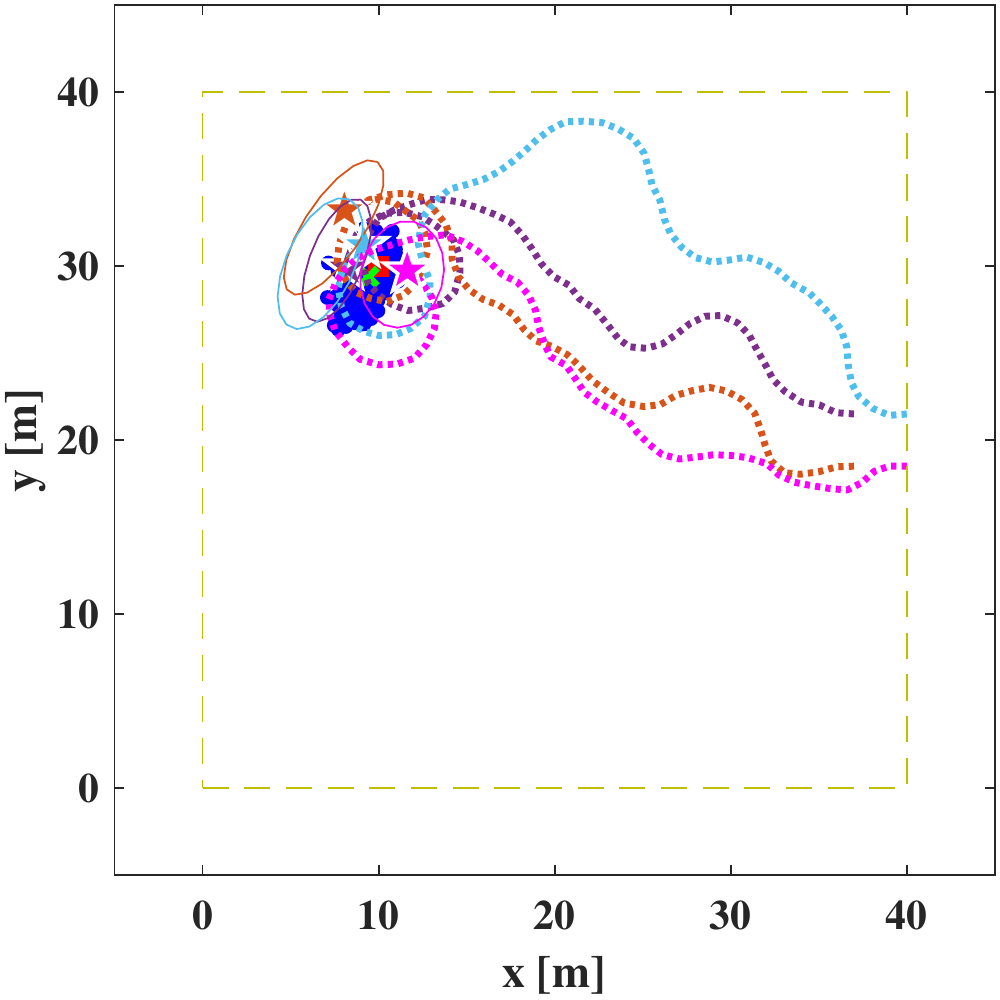}
		\caption{$t=55s$}
		\label{fig:2P_o55}
	\end{subfigure}
	\caption{Snapshots and resulting trajectories from (a)-(c) PF-only method, and (d)-(f) the proposed algorithm with noise covariance 2$P_0$; the red square, the scattered blue dots, and the green 'x' mark represent the true position, the particles, and the MMSE estimation of the target, respectively. The filled stars and the dotted lines indicate the true positions and trajectories of the agents, respectively. The empty stars show the estimated locations of the agents in PF-only method, while the empty ellipses indicate the distributions of the integrated estimations for the agents' states.}
	\label{fig:2P}
\end{figure}

Fig.~\ref{fig:2P} shows the simulation results for a single trial with the covariance matrix $2P_0$. Comparing the proposed algorithm and the PF-only algorithm, the most distinct difference is the path of each agent. In Fig.~\ref{fig:2P}(b) and \ref{fig:2P}(e), the four agents show the similar paths in both cases while the cumulative errors are relatively small. However, as time goes, the errors accumulate more and result in the different paths as shown in Fig.~\ref{fig:2P}(c) and \ref{fig:2P}(f). The PF-only algorithm does not take account of the motion noise and plans the next step movement for every single naive estimation. This causes the naive belief on the state of each agent deviates more from the true state as the noise accumulates. On the other hand, the proposed method plans the next step movement optimized for the overall distribution of the sensor state estimation. Also, it corrects the estimation based on the difference between the real and expected measurements. Thus, the proposed algorithm results in the paths of the agents converging near the target while PF-only method does not.

\begin{table}[b]\small
	\caption{\label{tab:table1} Mean and quartiles for estimation errors [m] at $t=100$ }
	\hspace{-0.3cm}
	\centering
	\begin{tabular}{c|ccc|ccc|ccc|ccc}
		\hline
		&      \multicolumn{6}{c|}{PF-only}& \multicolumn{6}{c}{Proposed}\\\cline{2-13}
		Noise&       &       Target&     &       &       Agent&     &
		&    Target&     &       &       Agent&\\\cline{2-13}
		[$\times P_0$]&  Q1& mean&   Q3& Q1& mean&   Q3&
		Q1& mean&   Q3& Q1& mean&   Q3\\\hline\hline
		0&      0.2850& 0.4479& 0.5827&    0.0000&   0.0000&   0.0000&
		0.2588& 0.4097& 0.5354&    0.0000& 0.0000&     0.0000\\
		0.5&  1.0150& 2.0351&     2.4301&    2.4380&   3.1485&   3.6132&
		0.7385& 1.4601&   2.0460&    1.1425&  1.7751&    2.1844\\
		1&   1.2636& 2.4706&    3.2346&    3.2924&   4.5308&   5.4962&
		1.0628& 2.1369&  3.0479&    1.6162&   2.6057&   3.2223\\
		2&    2.4417&      4.3329&    6.0445&    4.6507& 6.5438&     8.1971&
		1.5318&      3.0980&    4.4804&   2.0558&  3.7702&    4.8279\\
		4&  2.4240&      5.1430&    6.7322&    6.7592& 8.9025&     10.359&
		2.1903&      4.0769&    5.4416&   3.1977& 5.2704&     6.6684\\
		6&    3.6223&      6.5862&    8.1812&    8.5413& 11.115&     14.027&
		2.8424&      5.4246&    7.2331&   4.1324& 6.7993&     8.3706\\
		\hline
	\end{tabular}
\end{table}

For the performance in target estimation, the proposed algorithm shows better results, but the performance gap is not large. This small gap may come from two reasons; the fast convergence of particles and the counterbalance of the errors made by each other agent. As shown in Fig.~\ref{fig:2P}(b) and \ref{fig:2P}(e), the particles are already filtered close to the target while the accumulated noise of motion is small. Thus, the candidates are already picked well before affected by largely accumulated noises. Also, in Fig.~\ref{fig:2P}(c), the direction that each agent deviates from the target is not consistent. Then, we can expect that the errors made by the inaccurate estimations of the agents might cancel each other. 
Other typical results of each noise level are represented in Appendix.

\begin{figure}[h]
	\centering
	\begin{subfigure}{0.33\textwidth}
		\centering
		\includegraphics[width=\textwidth]{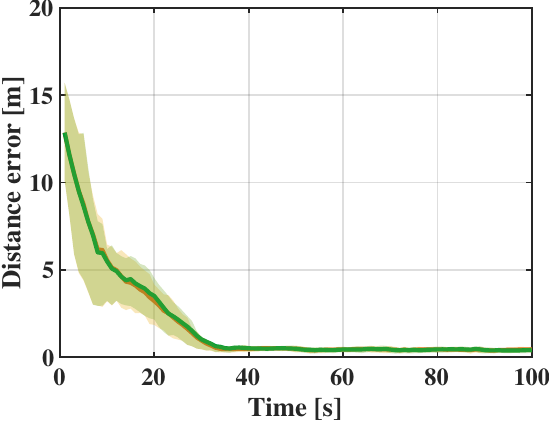}
		\caption{0}
		\label{fig:mc_target_0}
	\end{subfigure}
	\begin{subfigure}{0.33\textwidth}
		\centering
		\includegraphics[width=\textwidth]{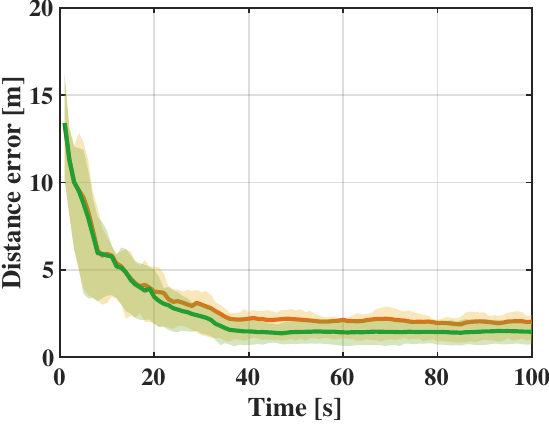}
		\caption{0.5$P_0$}
		\label{fig:mc_target_0.5}
	\end{subfigure}
	\begin{subfigure}{0.33\textwidth}
		\centering
		\includegraphics[width=\textwidth]{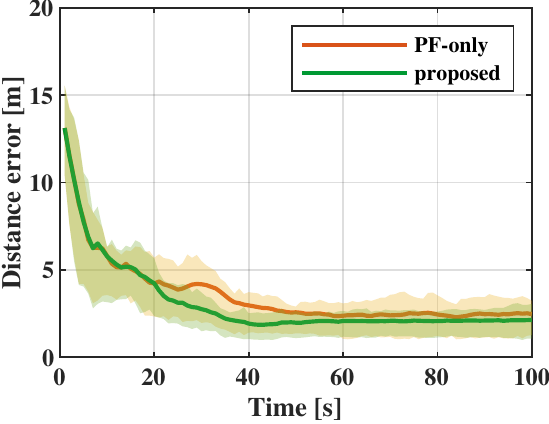}
		\caption{$P_0$}
		\label{fig:mc_target_1}
	\end{subfigure}
	
	\begin{subfigure}{0.33\textwidth}
		\centering
		\includegraphics[width=\textwidth]{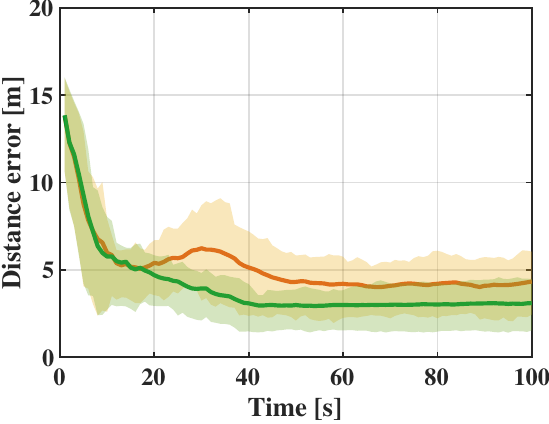}
		\caption{2$P_0$}
		\label{fig:mc_target_2}
	\end{subfigure}
	\begin{subfigure}{0.33\textwidth}
		\centering
		\includegraphics[width=\textwidth]{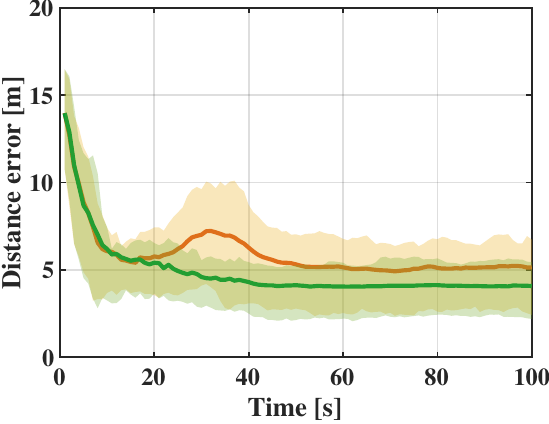}
		\caption{4$P_0$}
		\label{fig:mc_target_4}
	\end{subfigure}
	\begin{subfigure}{0.33\textwidth}
		\centering
		\includegraphics[width=\textwidth]{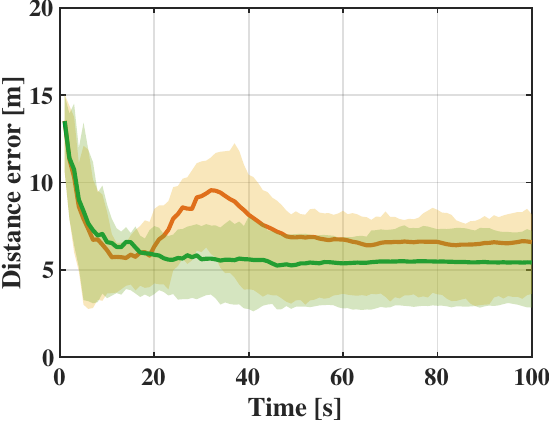}
		\caption{6$P_0$}
		\label{fig:mc_target_6}
	\end{subfigure}
	\caption{Change of localization errors of the target along time at each noise level. The solid lines indicate the mean values of the errors for 100 trials. The shaded regions represent the intervals between the $1^{st}$ and $3^{rd}$ quartiles.}
	\label{fig:mc_target}
%\end{figure}
%\begin{figure}[bt!]
	\vspace{0.2cm}
	\centering
	\begin{subfigure}{0.33\textwidth}
		\centering
		\includegraphics[width=\textwidth]{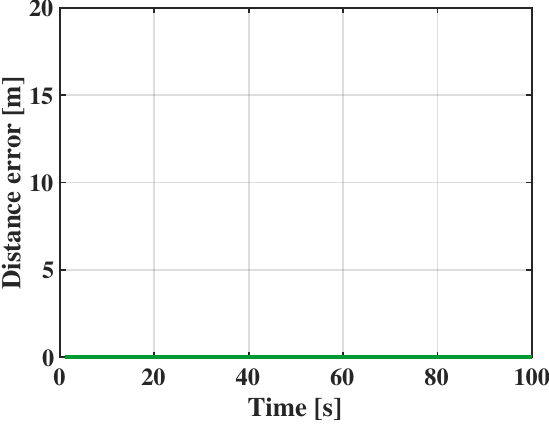}
		\caption{0}
		\label{fig:mc_agent_0}
	\end{subfigure}
	\begin{subfigure}{0.33\textwidth}
		\centering
		\includegraphics[width=\textwidth]{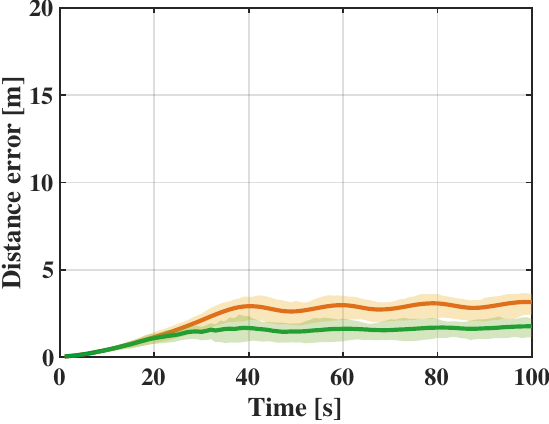}
		\caption{0.5$P_0$}
		\label{fig:mc_agent_0.5}
	\end{subfigure}
	\begin{subfigure}{0.33\textwidth}
		\centering
		\includegraphics[width=\textwidth]{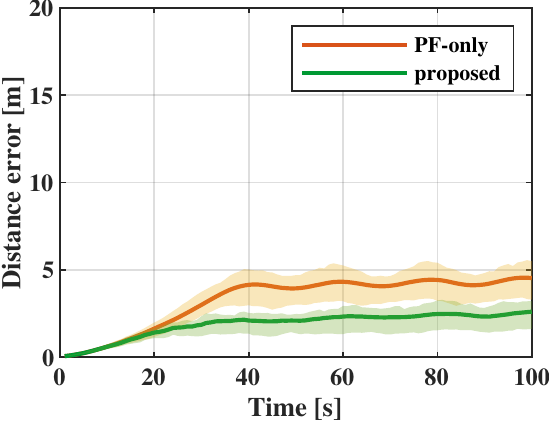}
		\caption{$P_0$}
		\label{fig:mc_agent_1}
	\end{subfigure}
	
	\begin{subfigure}{0.33\textwidth}
		\centering
		\includegraphics[width=\textwidth]{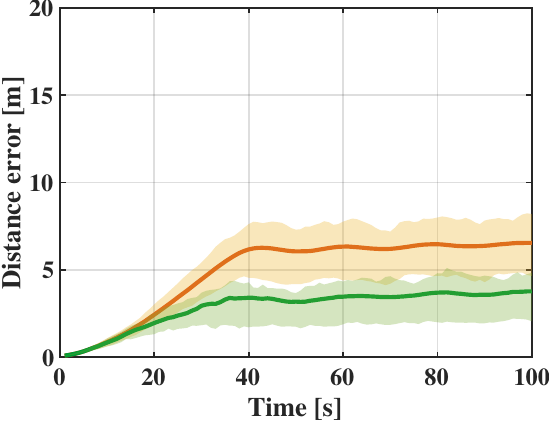}
		\caption{2$P_0$}
		\label{fig:mc_agent_2}
	\end{subfigure}
	\begin{subfigure}{0.33\textwidth}
		\centering
		\includegraphics[width=\textwidth]{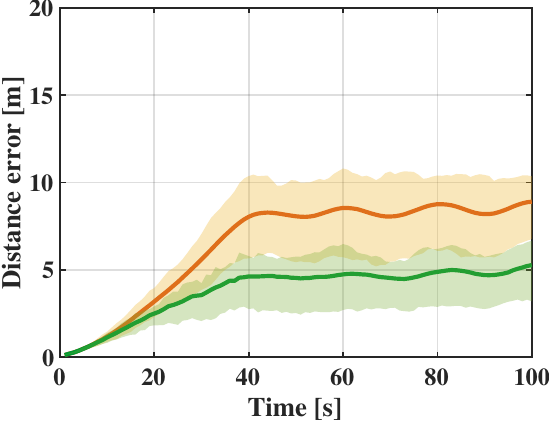}
		\caption{4$P_0$}
		\label{fig:mc_agent_4}
	\end{subfigure}
	\begin{subfigure}{0.33\textwidth}
		\centering
		\includegraphics[width=\textwidth]{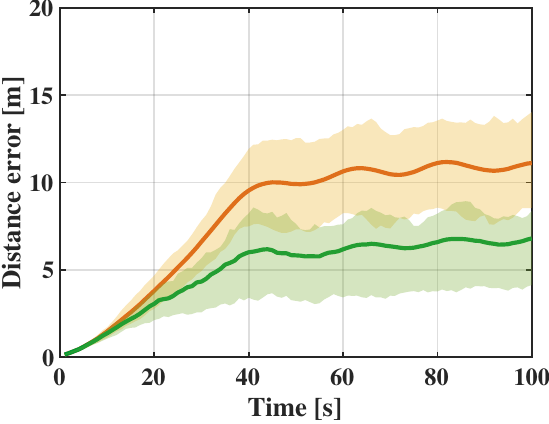}
		\caption{6$P_0$}
		\label{fig:mc_agent_6}
	\end{subfigure}
	\caption{Change of the average localization errors of the 4 agents along time at each noise level.}
	\label{fig:mc_agent}
\end{figure}

The overall results of Monte Carlo experiments are shown in Fig.~\ref{fig:mc_target} and \ref{fig:mc_agent}. When the noise is set to zero, both algorithms show the consistent localization performances as in Fig.~\ref{fig:mc_target}(a). As the noise level increases, the average error also increases, which is the expected behavior for the harder mission. In the same spirit of the analysis on Fig.~\ref{fig:2P}, the proposed algorithm shows better target localization performance in general, but the performance gap is not large. Meanwhile, the capability of estimating the agents' state is remarkably better for the proposed algorithm as shown in Fig.~\ref{fig:mc_agent}. This accurate sensor state estimation is beneficial to further missions after localizing the target such as moving the target. Table~\ref{tab:table1} shows the numerical values of the results at $t=100$.

\clearpage

\section{Conclusion}

In this paper, we have presented a new algorithm that reflects the uncertainty of the states of the sensing agents in planning sensor networks. For this purpose, we have combined two Bayesian filters and utilized them to evaluate the mutual information between the future sensor measurements and the target state. This approach does not necessitate the extra sensors nor the prior knowledge of the environments. The simulation results in the GPS-denied scenarios have shown that this approach increases the accuracy of estimating the states of both the target and the sensor network. In the future, the proposed method can be extended to non-myopic planning and/or planning under complex environments such as one including some obstacles.

%\clearpage

\vspace{2cm}

\section*{Appendix A. Simulation Results with Different Noise Levels}

\vspace{1cm}

\begin{figure}[hbt!]
% \begin{minipage}[c][\textheight]{\textwidth}
     \centering
     \begin{subfigure}[b]{.24\textwidth}
         \centering
         \includegraphics[width=\textwidth]{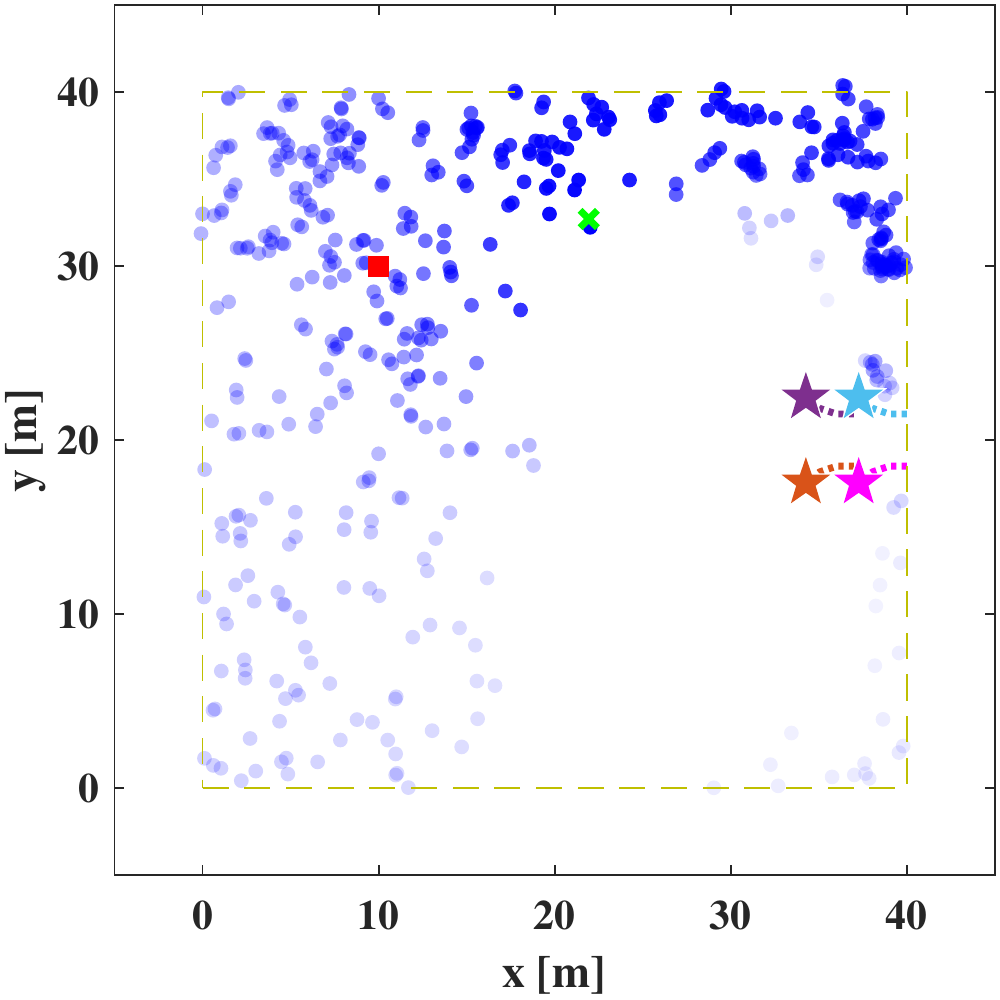}
         \caption{$t=3s$}
     \end{subfigure}
     \begin{subfigure}[b]{.24\textwidth}
         \centering
         \includegraphics[width=\textwidth]{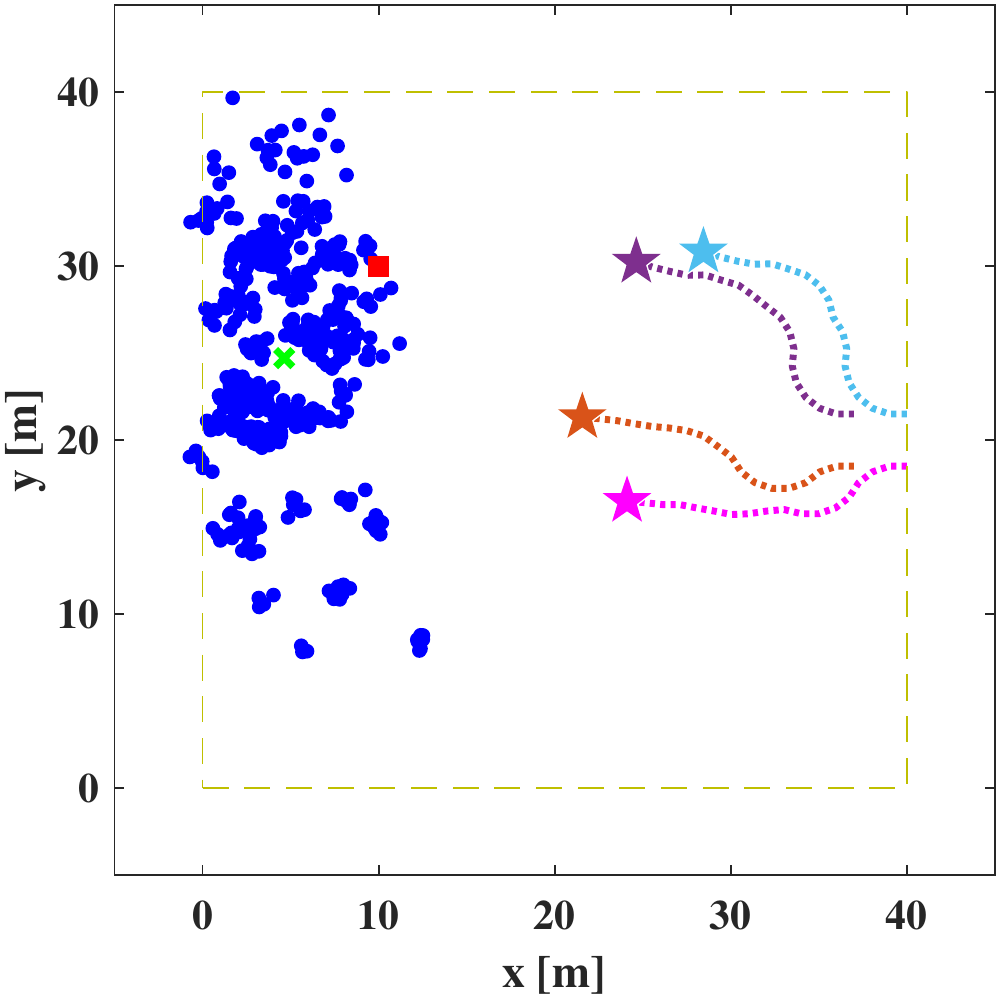}
         \caption{$t=17s$}
     \end{subfigure}
     \begin{subfigure}[b]{.24\textwidth}
         \centering
         \includegraphics[width=\textwidth]{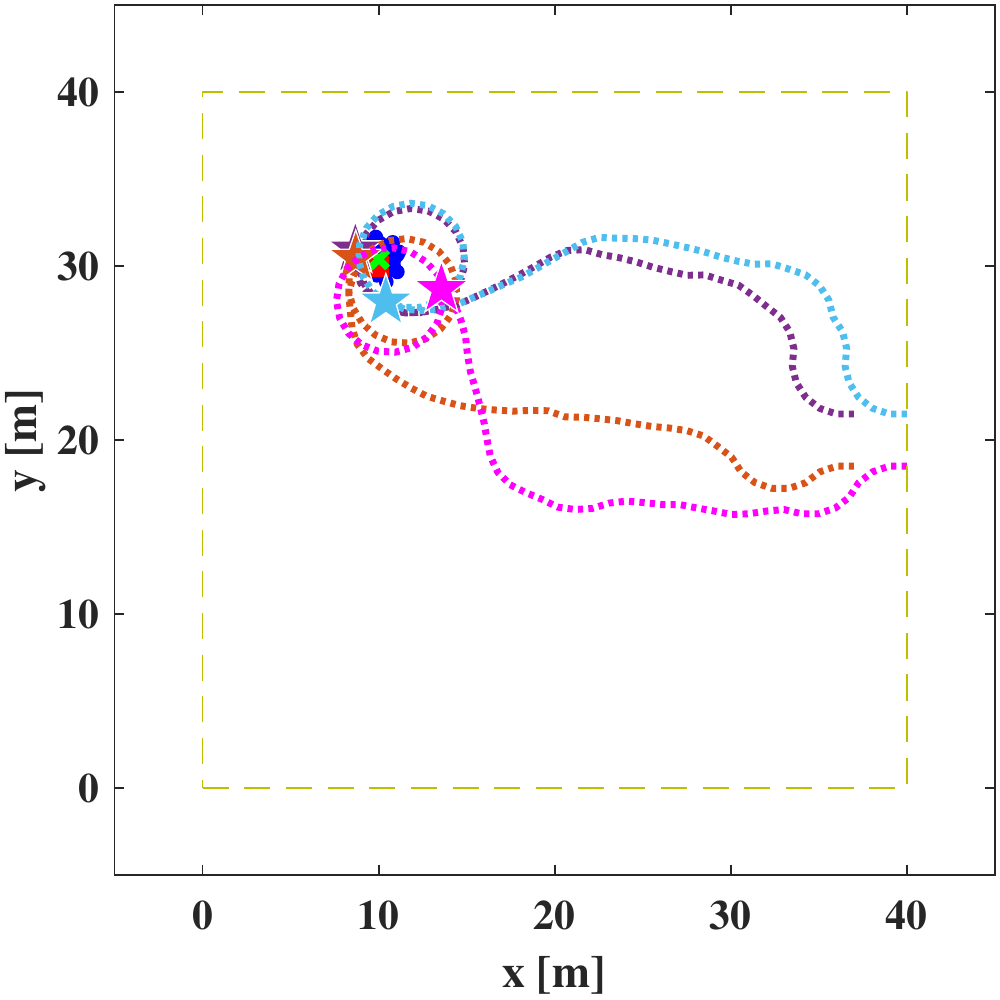}
         \caption{$t=55s$}
     \end{subfigure}\\
     \begin{subfigure}[b]{0.24\textwidth}
         \centering
         \includegraphics[width=\textwidth]{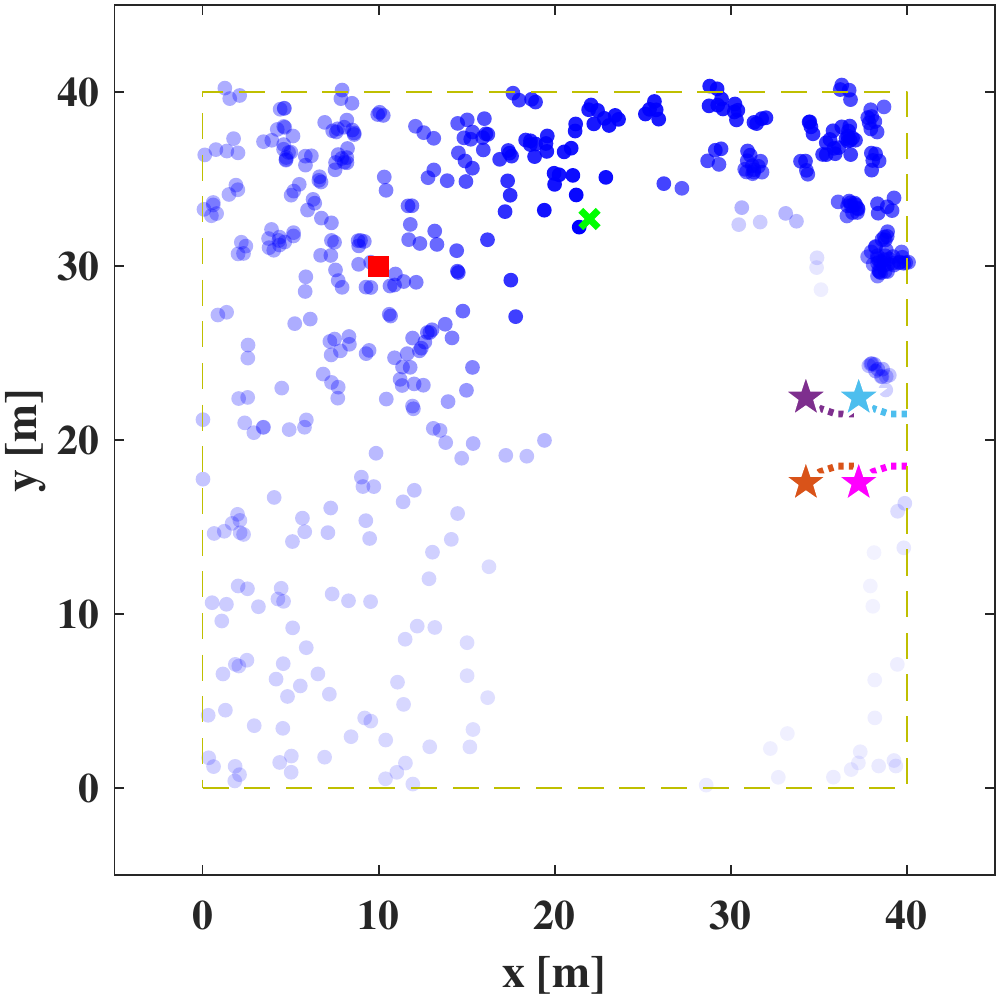}
         \caption{$t=3s$}
     \end{subfigure}
     \begin{subfigure}[b]{.24\textwidth}
         \centering
         \includegraphics[width=\textwidth]{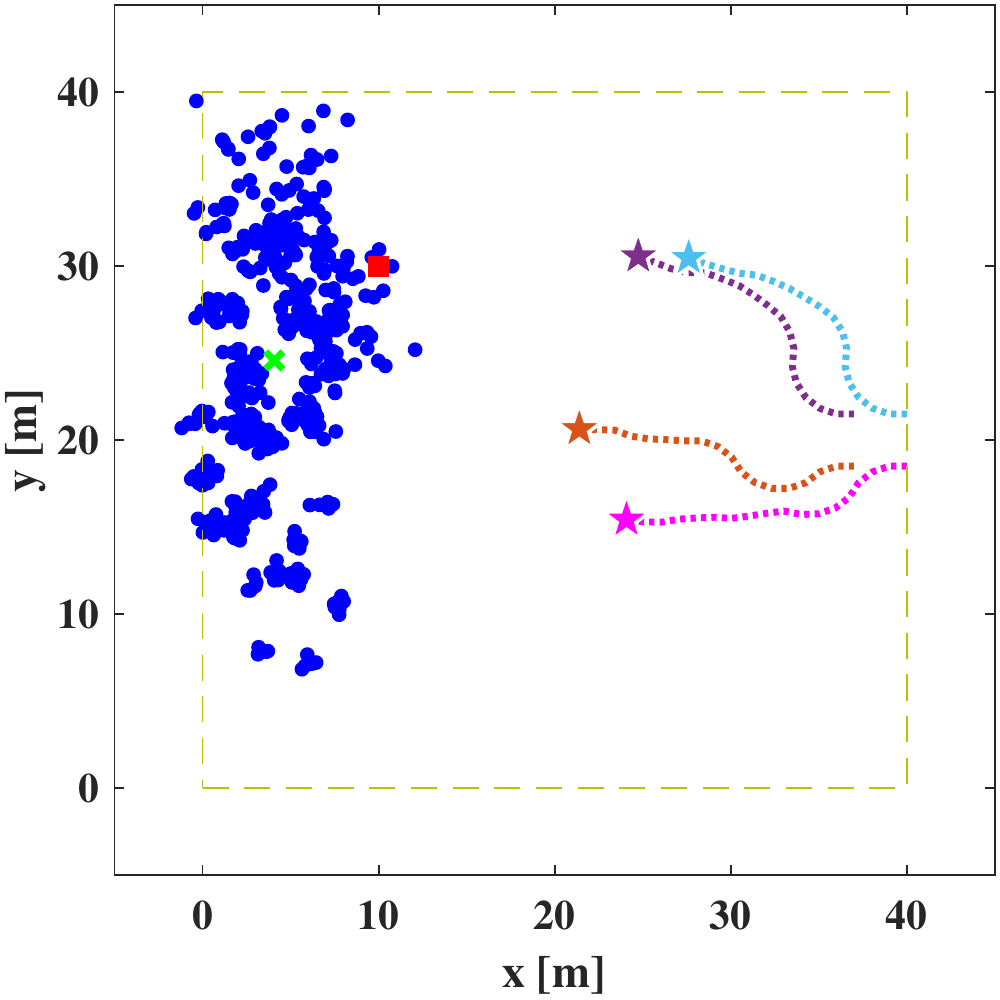}
         \caption{$t=17s$}
     \end{subfigure}
     \begin{subfigure}[b]{.24\textwidth}
         \centering
         \includegraphics[width=\textwidth]{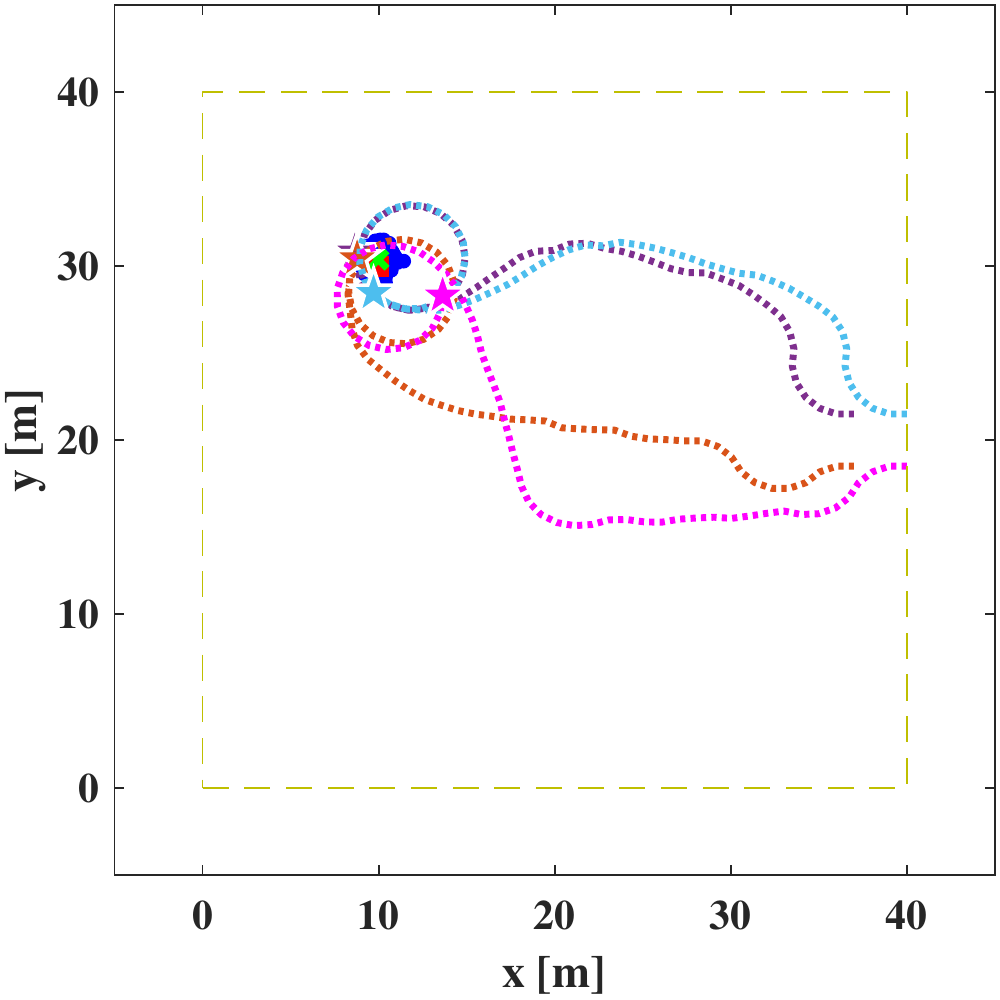}
         \caption{$t=55s$}
     \end{subfigure}
    \caption{Snapshots and resulting trajectories from (a)-(c) PF-only method, and (d)-(f) the proposed algorithm with zero noise}
    \label{fig:0P}
% \end{minipage}
\end{figure}

\begin{figure}[hbt!]
% \begin{minipage}[c][\textheight]{\textwidth}
     \centering
     \begin{subfigure}[b]{.24\textwidth}
         \centering
         \includegraphics[width=\textwidth]{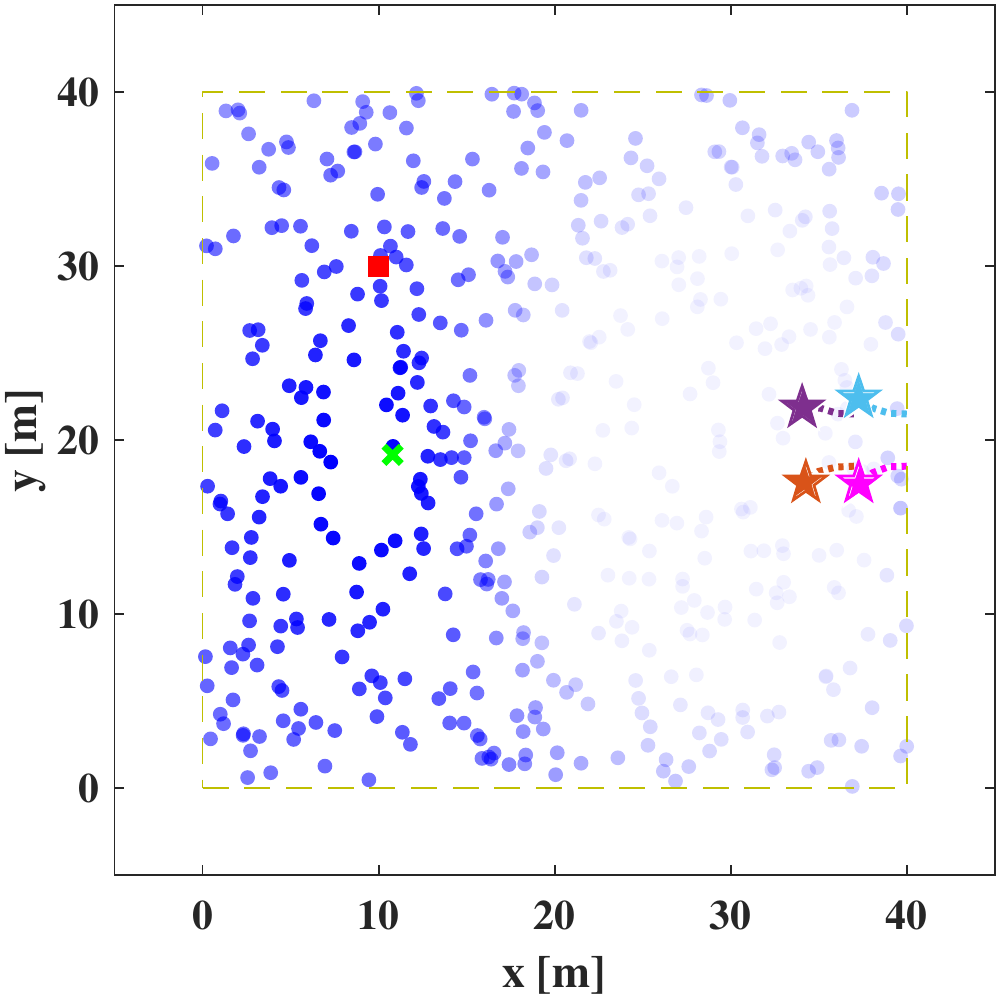}
         \caption{$t=3s$}
     \end{subfigure}
     \begin{subfigure}[b]{.24\textwidth}
         \centering
         \includegraphics[width=\textwidth]{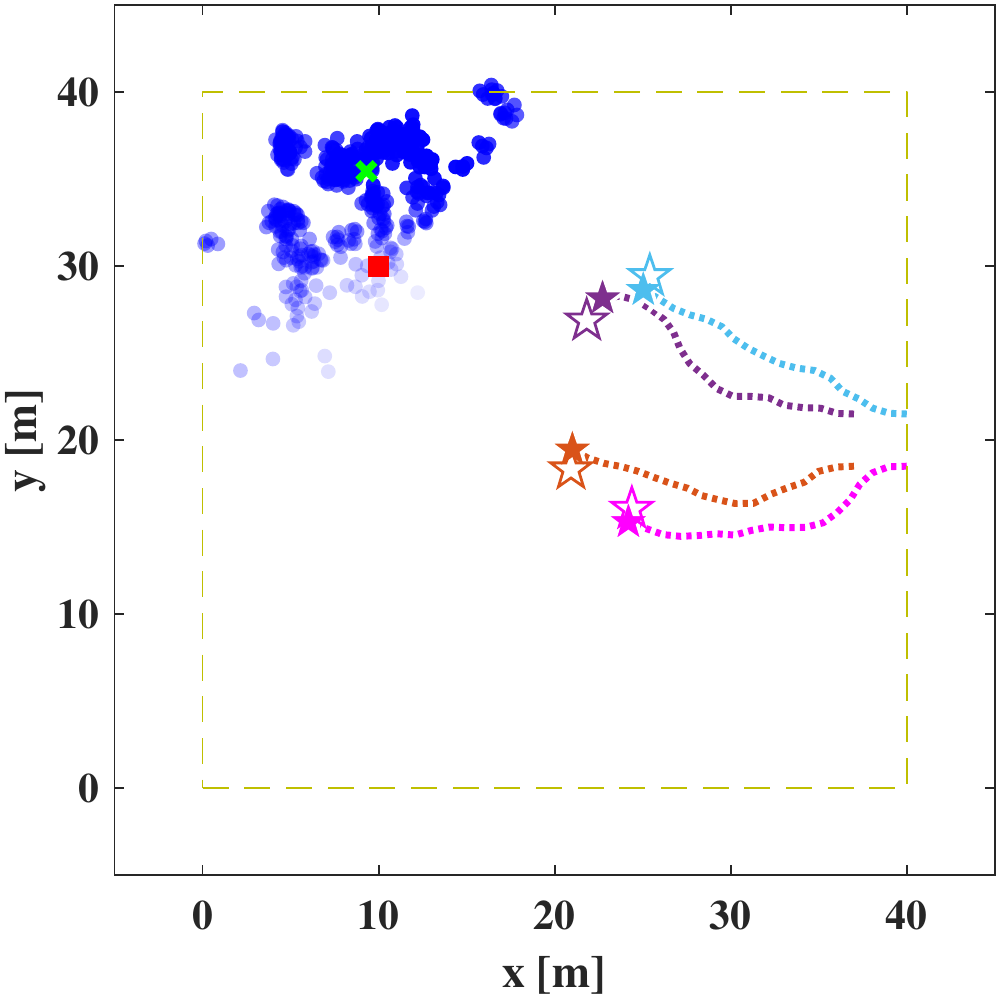}
         \caption{$t=17s$}
     \end{subfigure}
     \begin{subfigure}[b]{.24\textwidth}
         \centering
         \includegraphics[width=\textwidth]{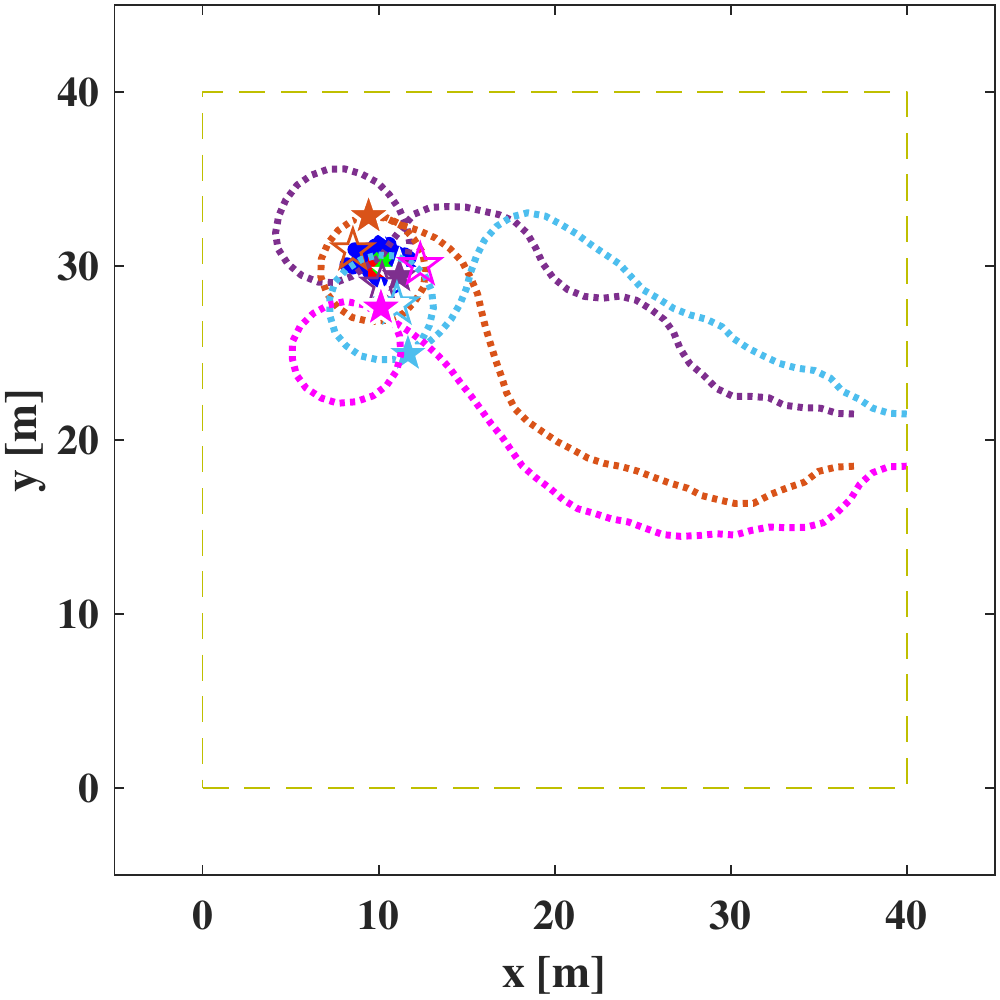}
         \caption{$t=55s$}
     \end{subfigure}\\
     \begin{subfigure}[b]{0.24\textwidth}
         \centering
         \includegraphics[width=\textwidth]{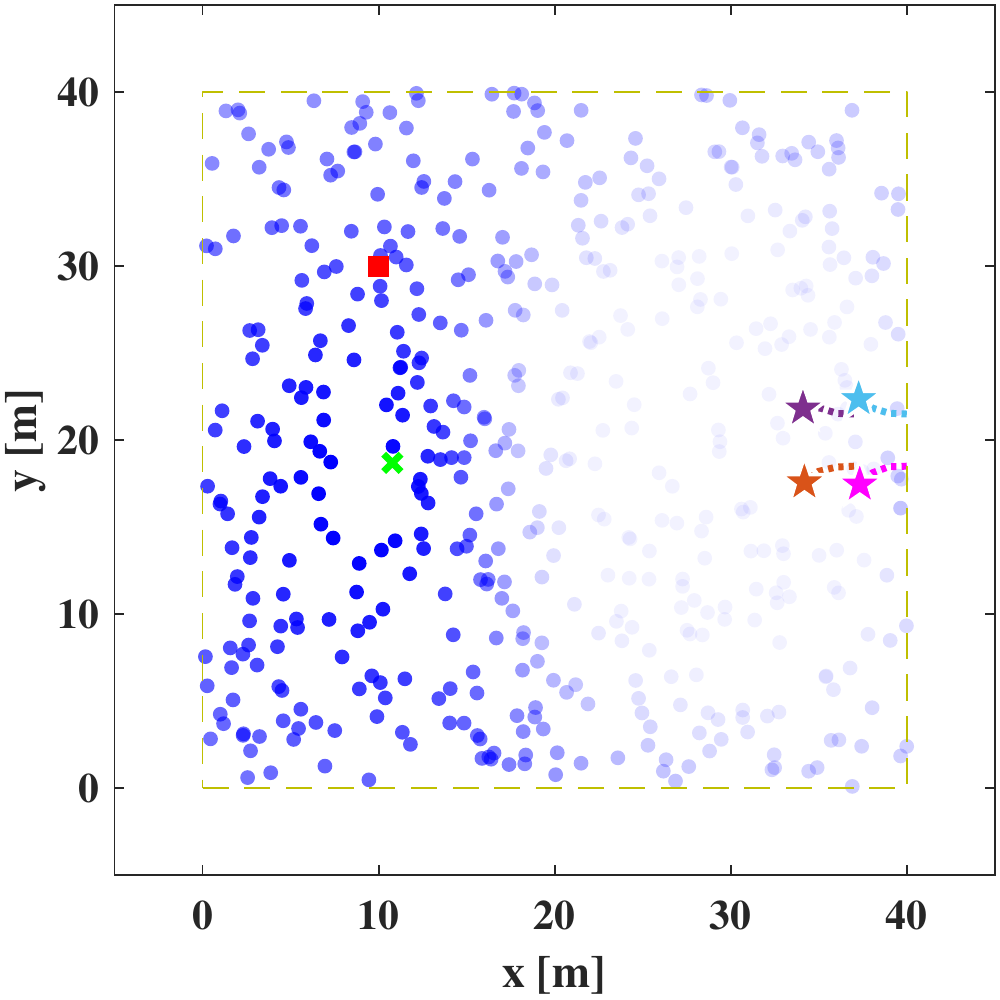}
         \caption{$t=3s$}
     \end{subfigure}
     \begin{subfigure}[b]{.24\textwidth}
         \centering
         \includegraphics[width=\textwidth]{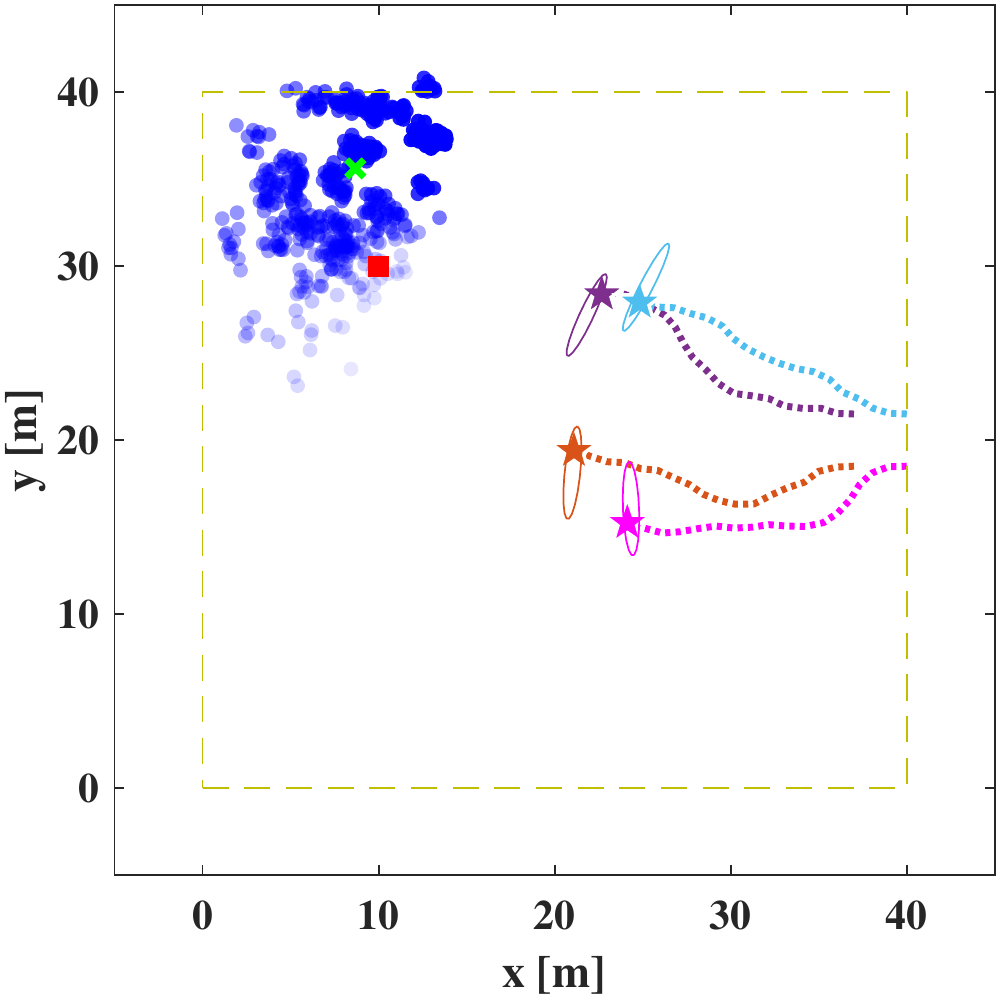}
         \caption{$t=17s$}
     \end{subfigure}
     \begin{subfigure}[b]{.24\textwidth}
         \centering
         \includegraphics[width=\textwidth]{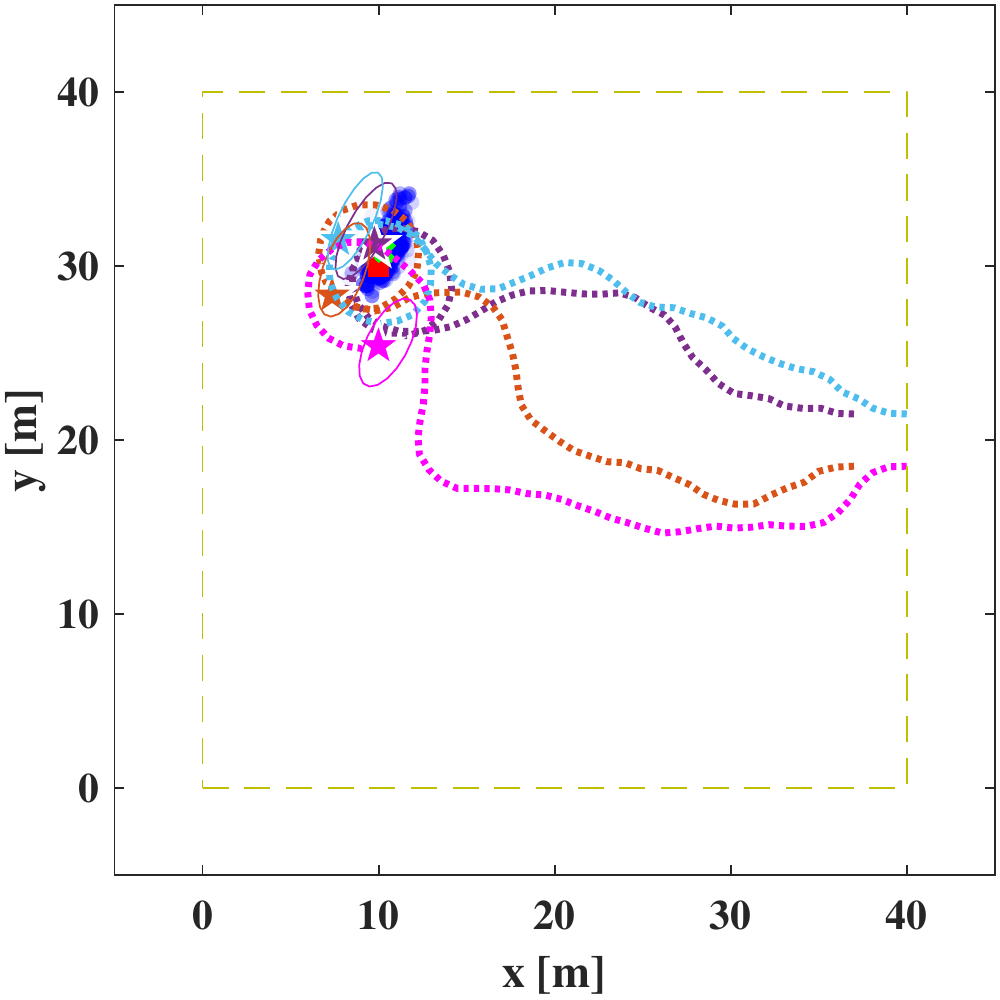}
         \caption{$t=55s$}
     \end{subfigure}
    \caption{Snapshots and resulting trajectories from (a)-(c) PF-only method, and (d)-(f) the proposed algorithm with noise covariance 0.5$P_0$}
    \label{fig:0.5P}
% \end{minipage}
\end{figure}

\begin{figure}[hbt!]
% \begin{minipage}[c][\textheight]{\textwidth}
     \centering
     \begin{subfigure}[b]{.24\textwidth}
         \centering
         \includegraphics[width=\textwidth]{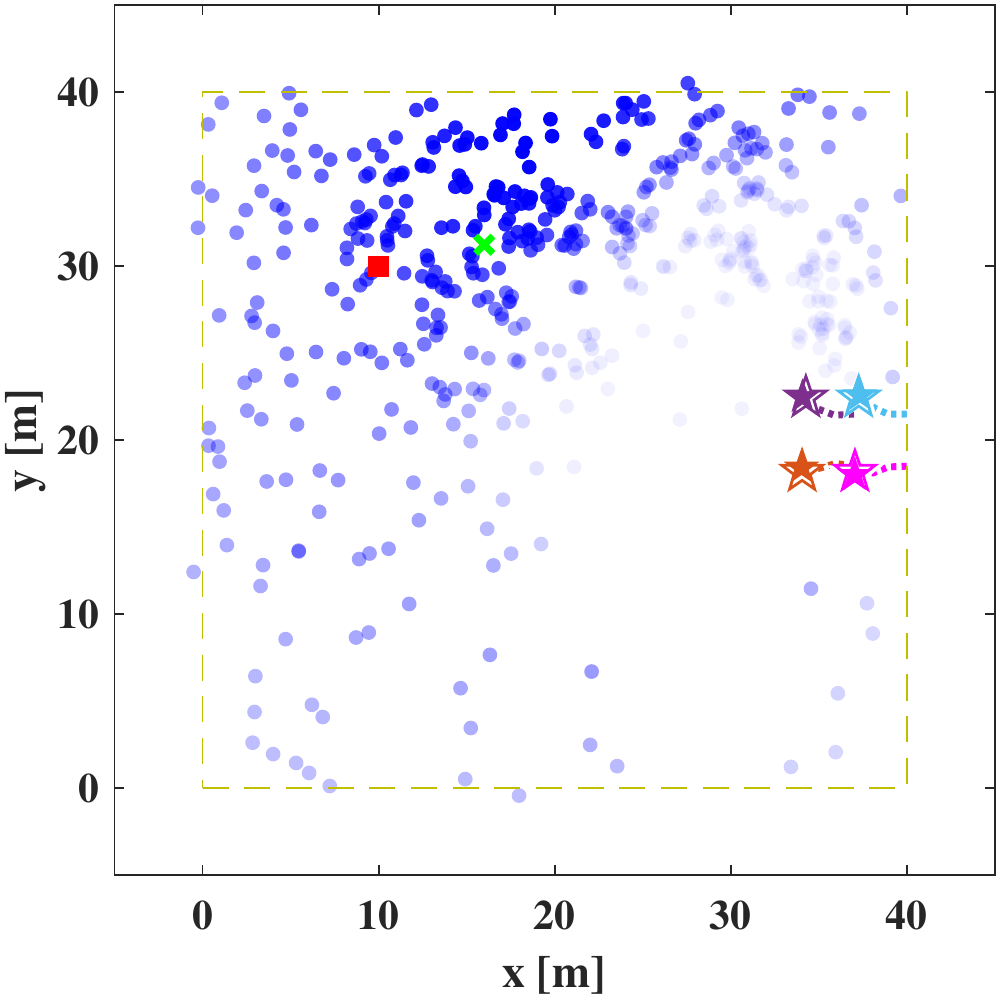}
         \caption{$t=3s$}
     \end{subfigure}
     \begin{subfigure}[b]{.24\textwidth}
         \centering
         \includegraphics[width=\textwidth]{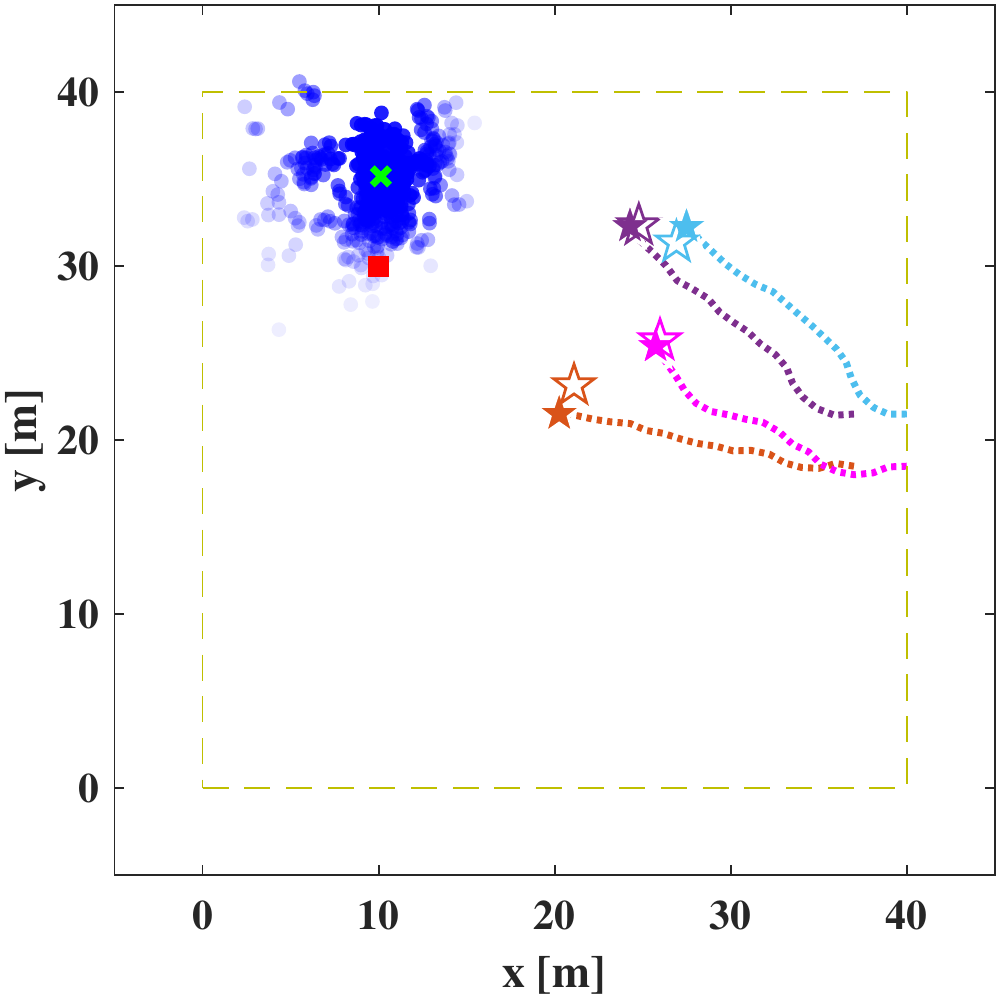}
         \caption{$t=17s$}
     \end{subfigure}
     \begin{subfigure}[b]{.24\textwidth}
         \centering
         \includegraphics[width=\textwidth]{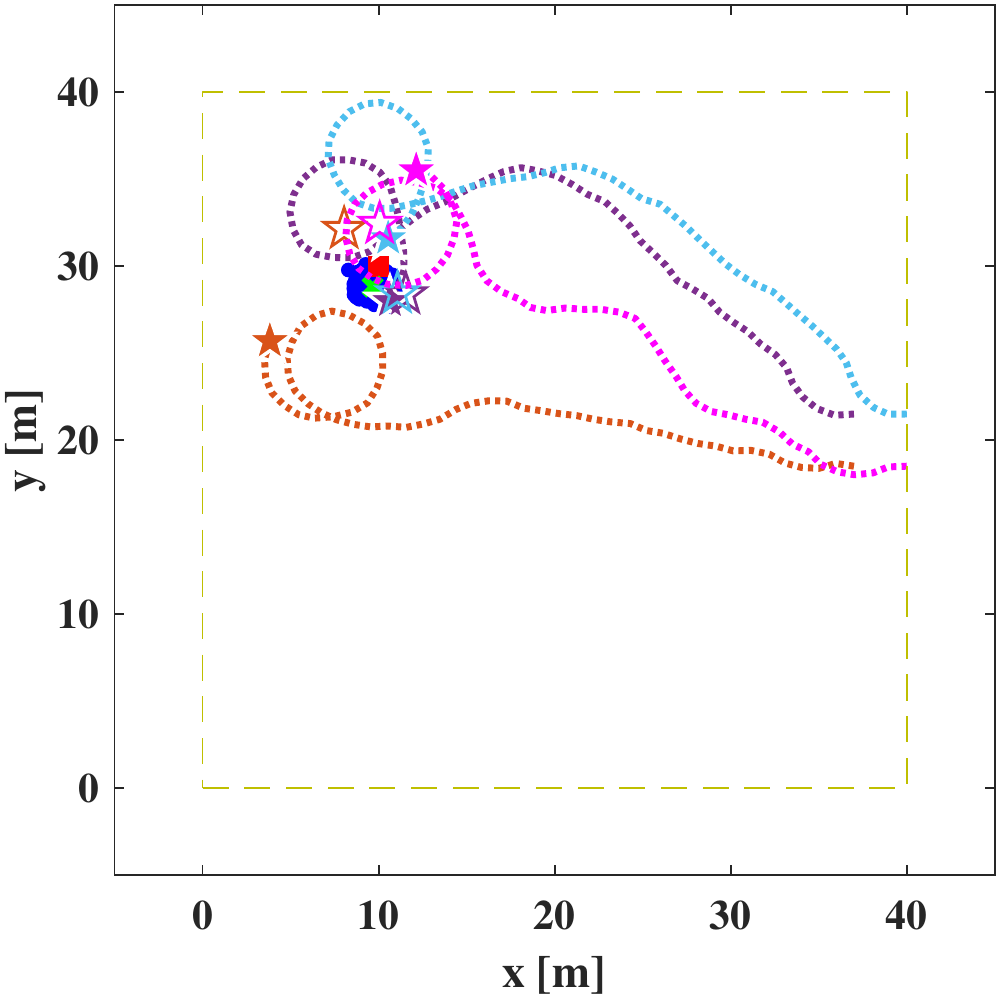}
         \caption{$t=55s$}
     \end{subfigure}\\
     \begin{subfigure}[b]{.24\textwidth}
         \centering
         \includegraphics[width=\textwidth]{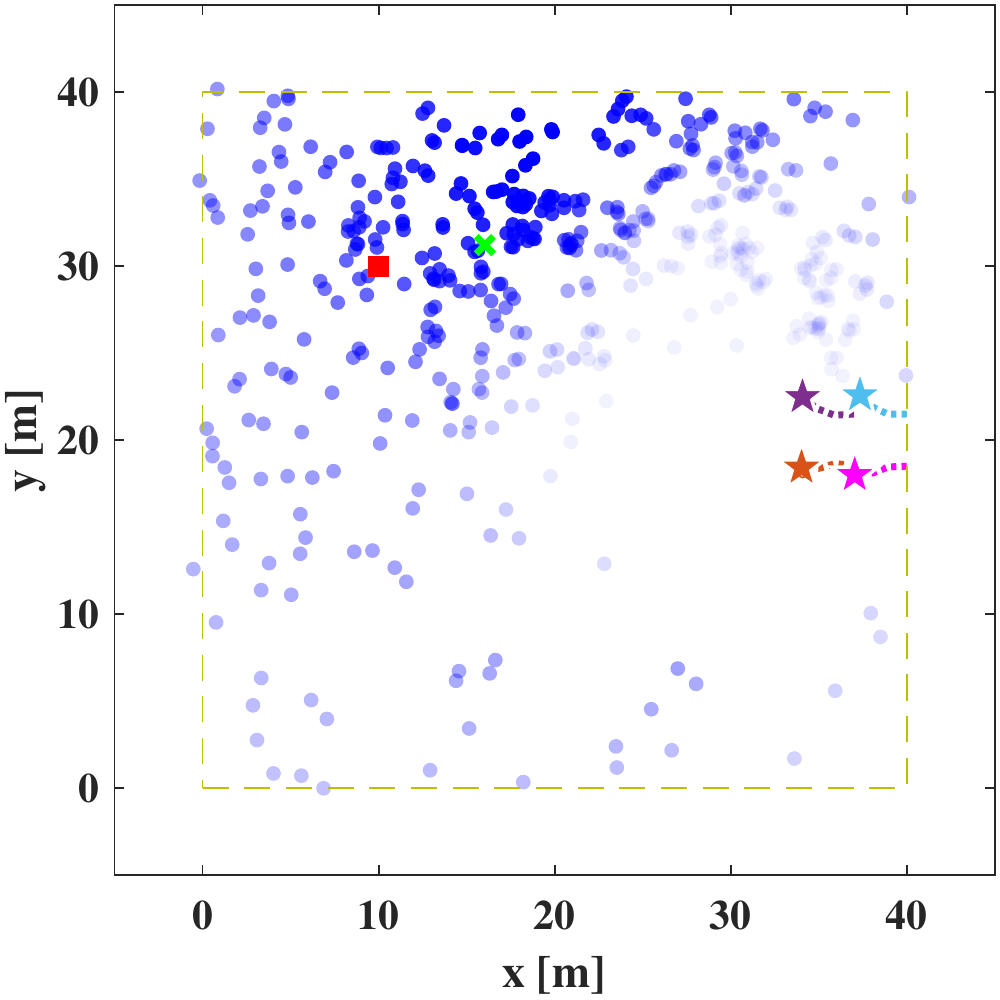}
         \caption{$t=3s$}
     \end{subfigure}
     \begin{subfigure}[b]{.24\textwidth}
         \centering
         \includegraphics[width=\textwidth]{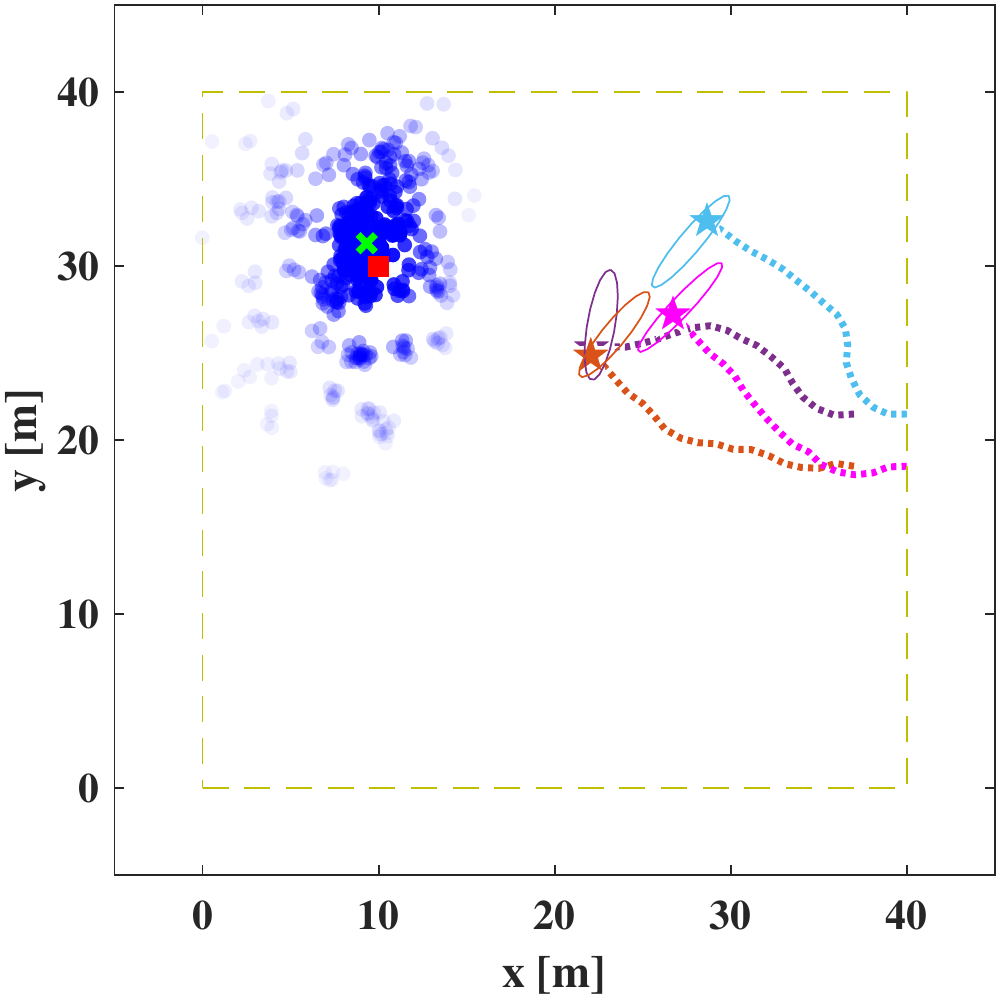}
         \caption{$t=17s$}
     \end{subfigure}
     \begin{subfigure}[b]{.24\textwidth}
         \centering
         \includegraphics[width=\textwidth]{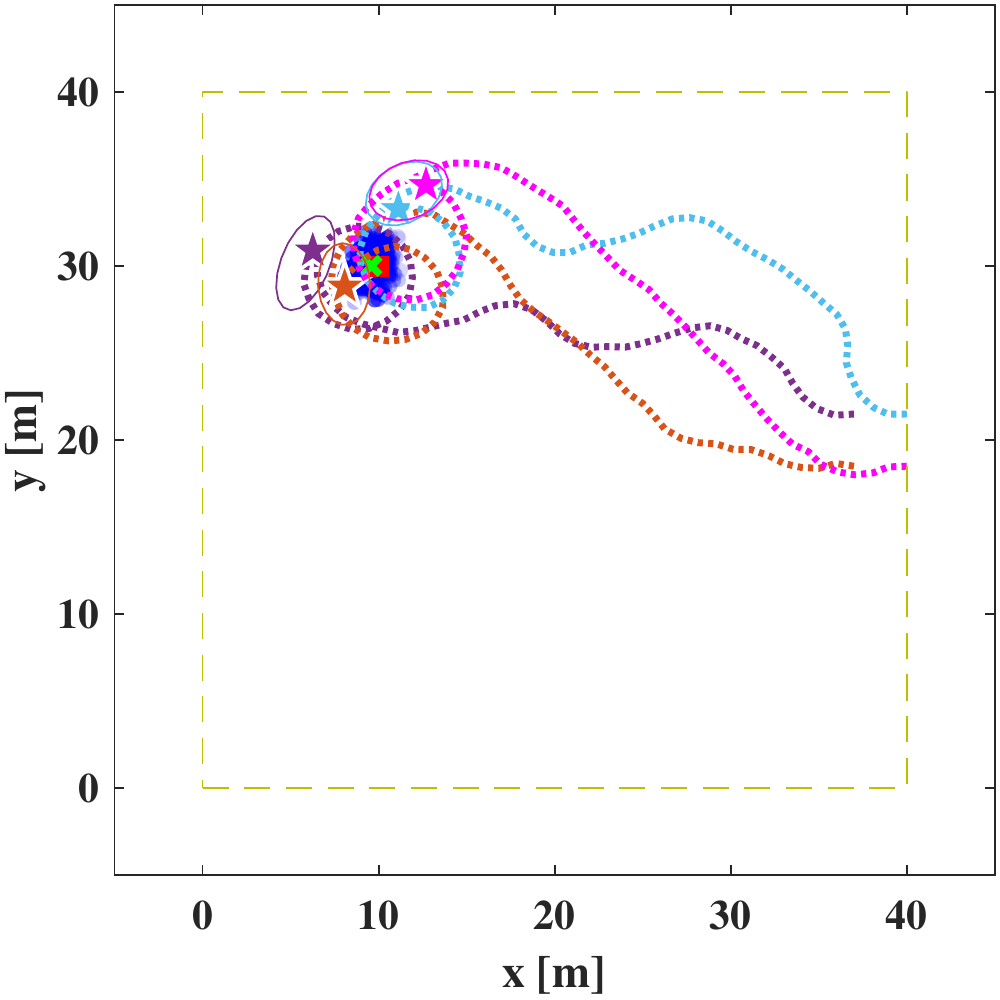}
         \caption{$t=55s$}
     \end{subfigure}
    \caption{Snapshots and resulting trajectories from (a)-(c) PF-only method, and (d)-(f) the proposed algorithm with noise covariance $P_0$}
    \label{fig:P}
% \end{minipage}
\end{figure}
~\\
\begin{figure}[hbt!]
% \begin{minipage}[c][\textheight]{\textwidth}
     \centering
     \begin{subfigure}[b]{.24\textwidth}
         \centering
         \includegraphics[width=\textwidth]{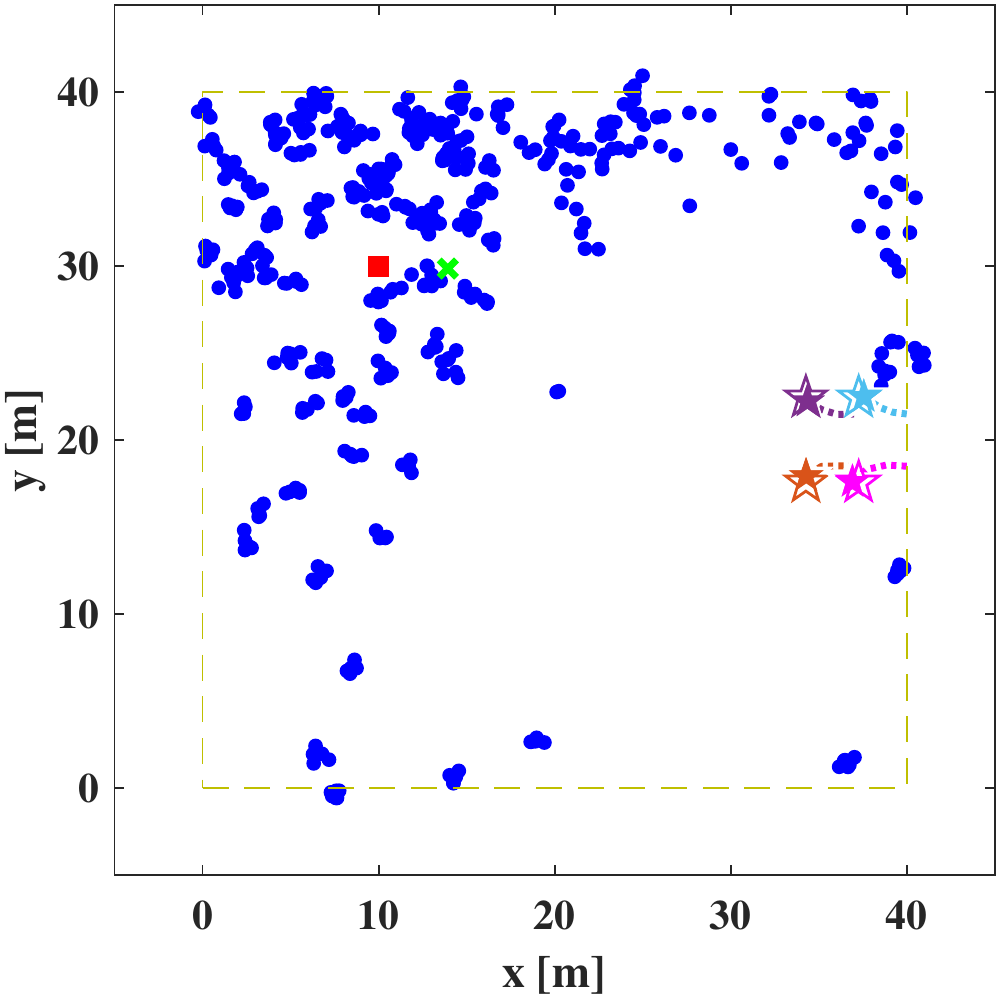}
         \caption{$t=3s$}
     \end{subfigure}
     \begin{subfigure}[b]{.24\textwidth}
         \centering
         \includegraphics[width=\textwidth]{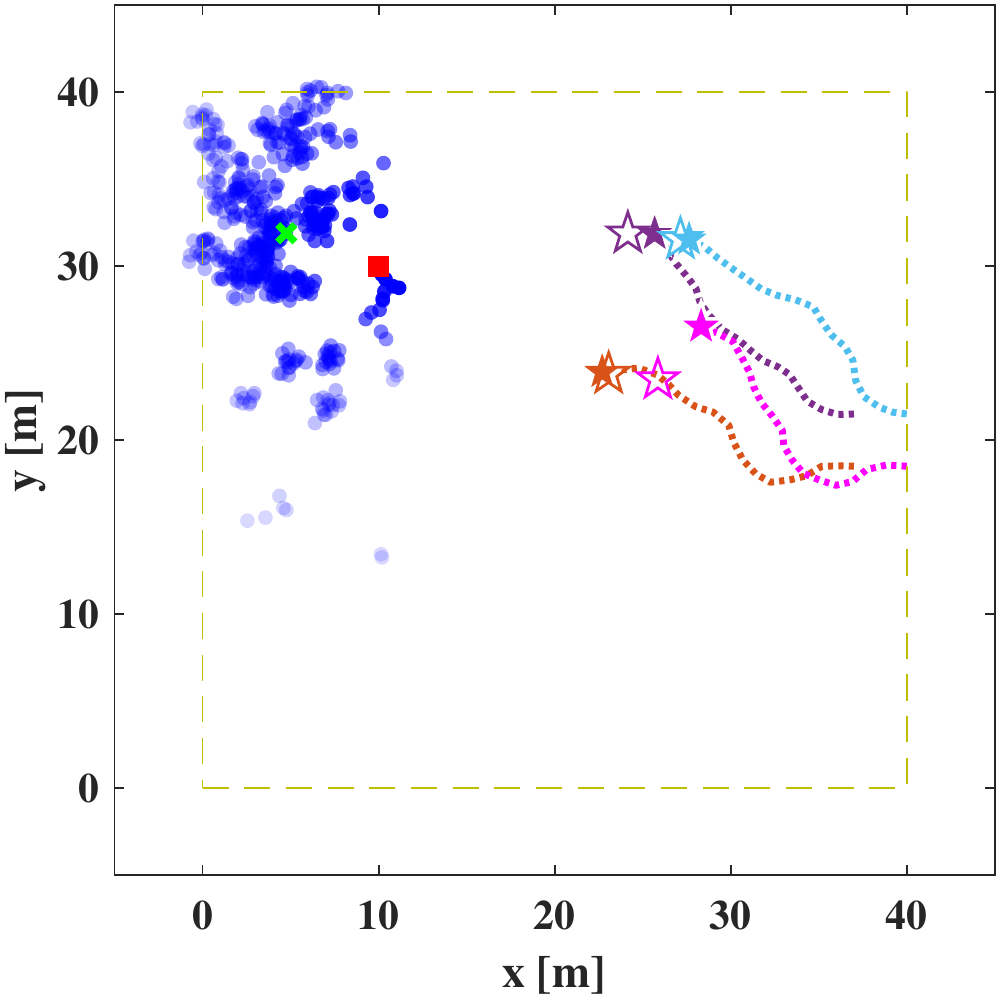}
         \caption{$t=17s$}
     \end{subfigure}
     \begin{subfigure}[b]{.24\textwidth}
         \centering
         \includegraphics[width=\textwidth]{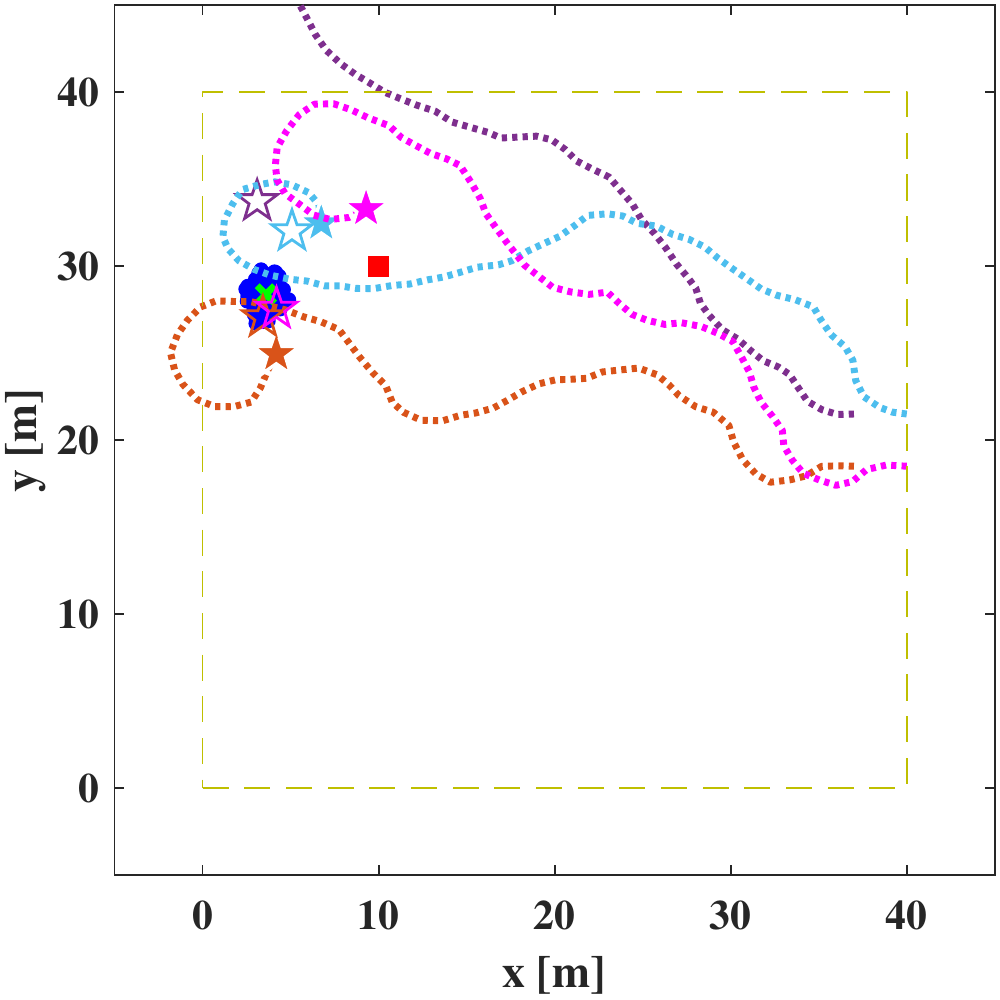}
         \caption{$t=55s$}
     \end{subfigure}\\
     \begin{subfigure}[b]{.24\textwidth}
         \centering
         \includegraphics[width=\textwidth]{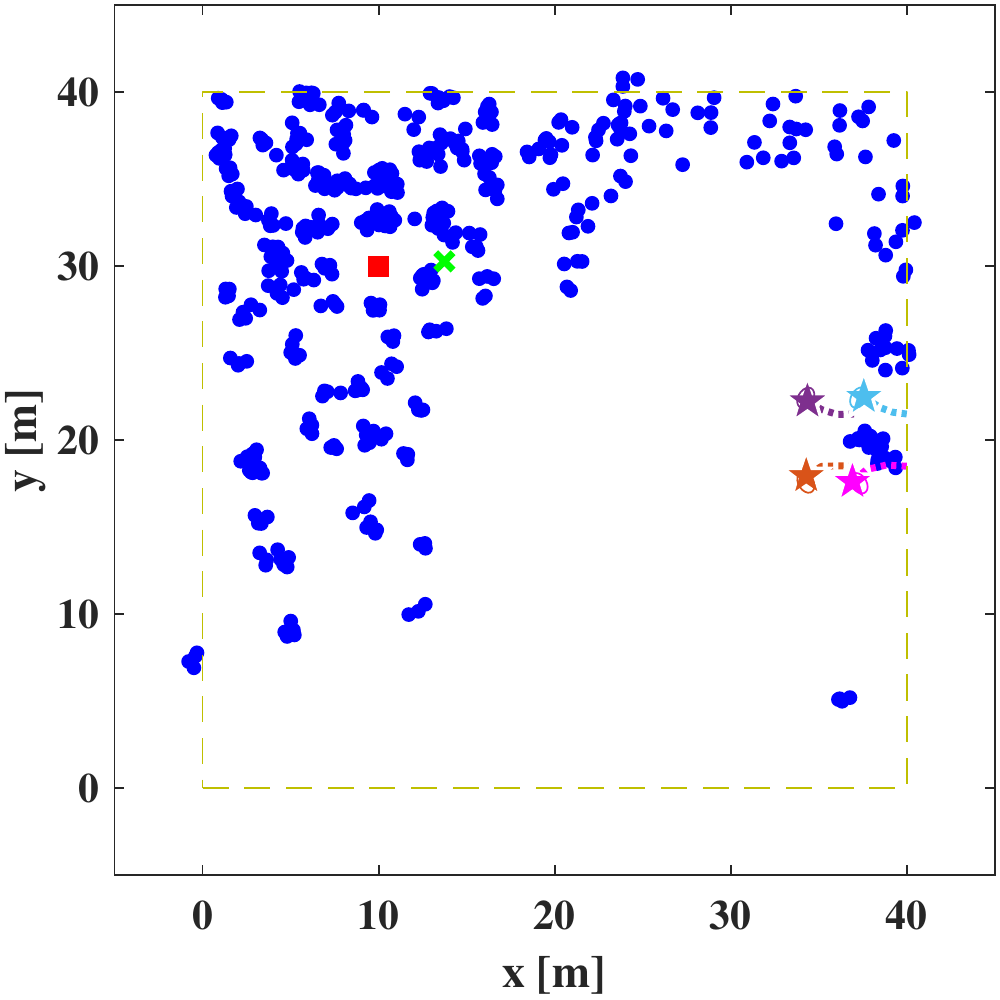}
         \caption{$t=3s$}
     \end{subfigure}
     \begin{subfigure}[b]{.24\textwidth}
         \centering
         \includegraphics[width=\textwidth]{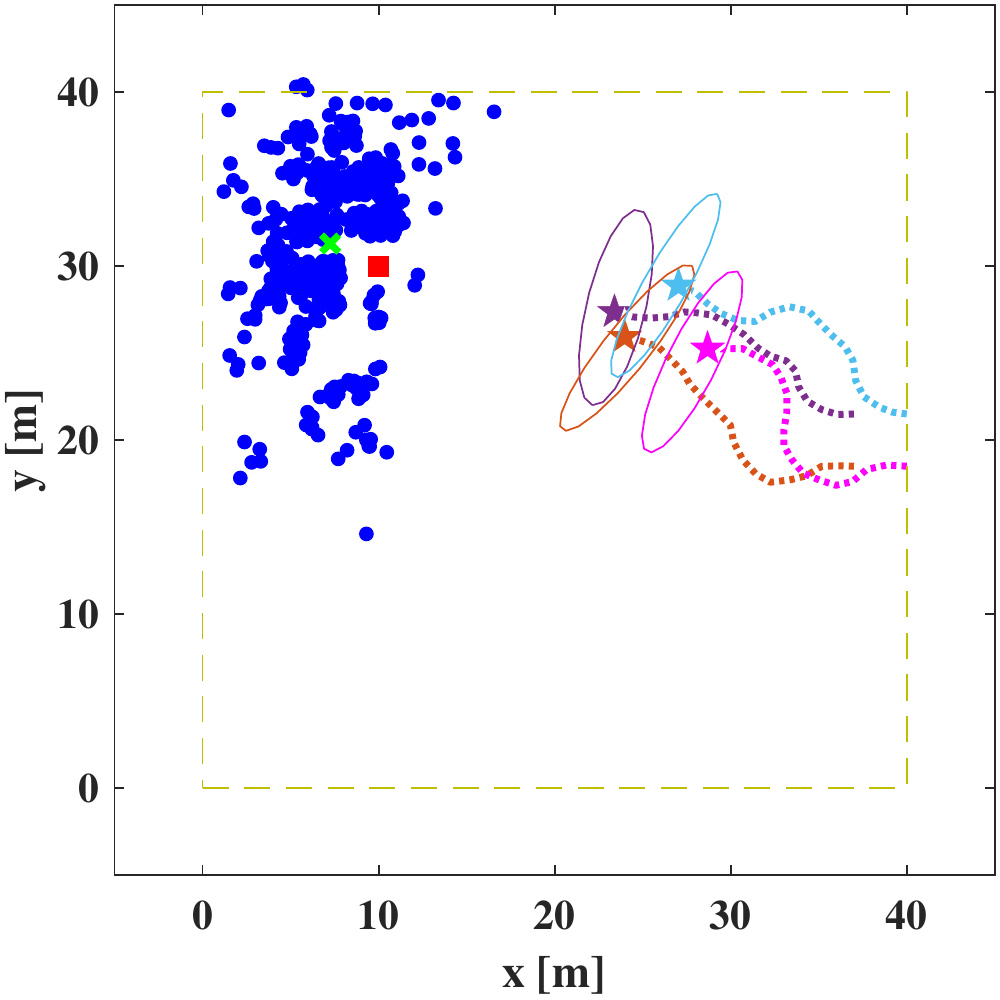}
         \caption{$t=17s$}
     \end{subfigure}
     \begin{subfigure}[b]{.24\textwidth}
         \centering
         \includegraphics[width=\textwidth]{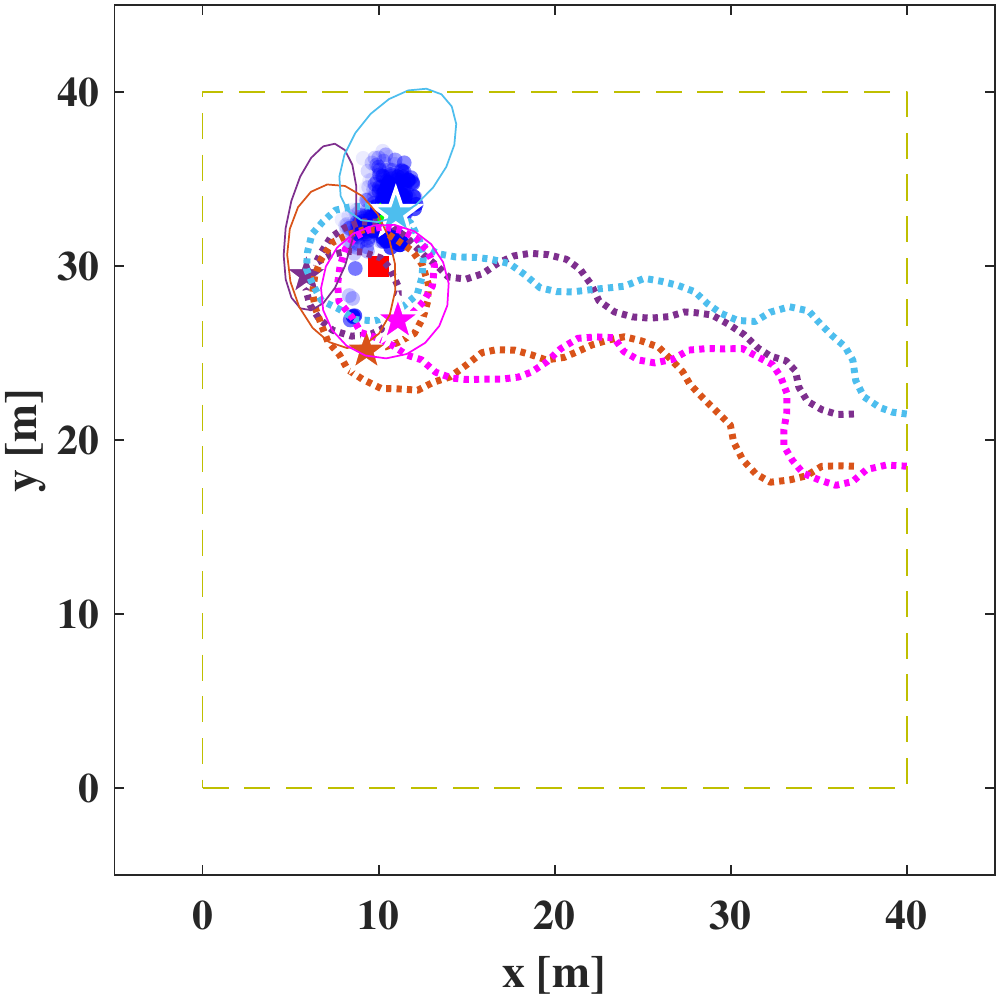}
         \caption{$t=55s$}
     \end{subfigure}
    \caption{Snapshots and resulting trajectories from (a)-(c) PF-only method, and (d)-(f) the proposed algorithm with noise covariance 4$P_0$}
    \label{fig:4P}
% \end{minipage}
\end{figure}

\begin{figure}[hbt!]
% \begin{minipage}[c][\textheight]{\textwidth}
     \centering
     \begin{subfigure}[b]{.24\textwidth}
         \centering
         \includegraphics[width=\textwidth]{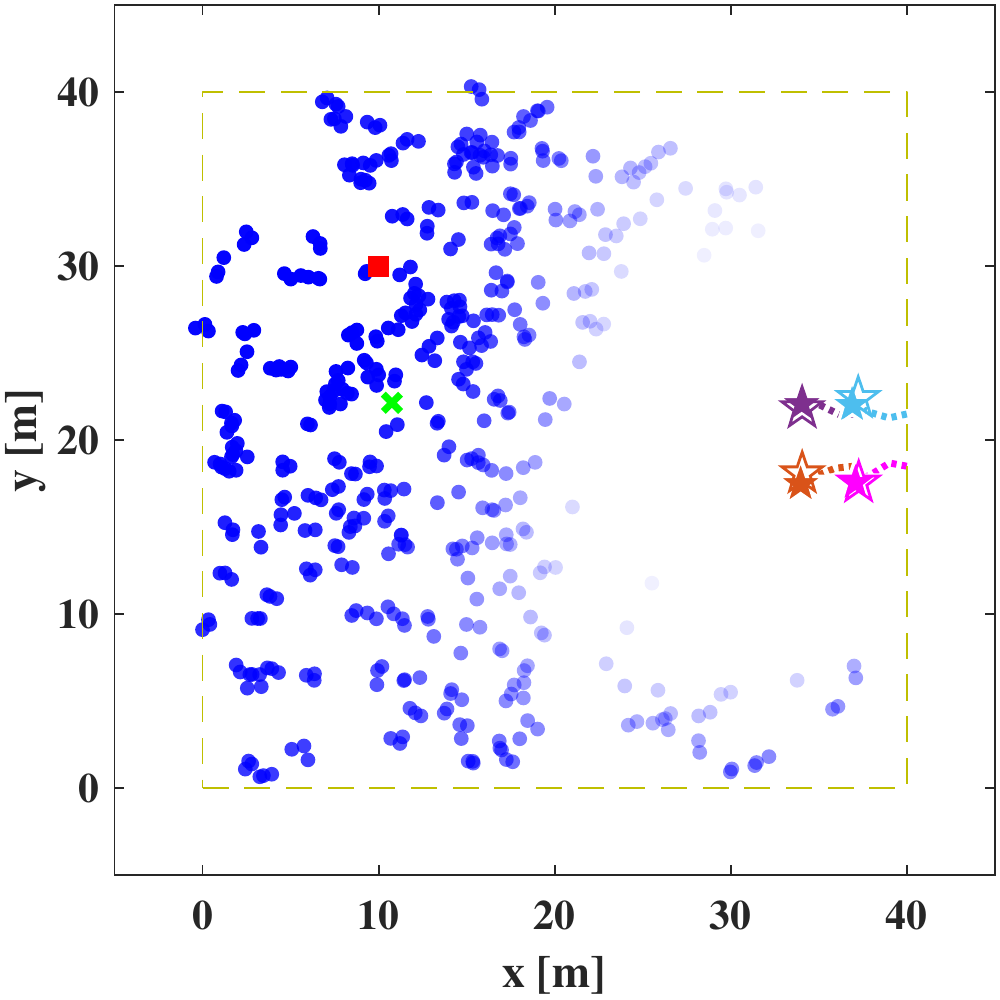}
         \caption{$t=3s$}
     \end{subfigure}
     \begin{subfigure}[b]{.24\textwidth}
         \centering
         \includegraphics[width=\textwidth]{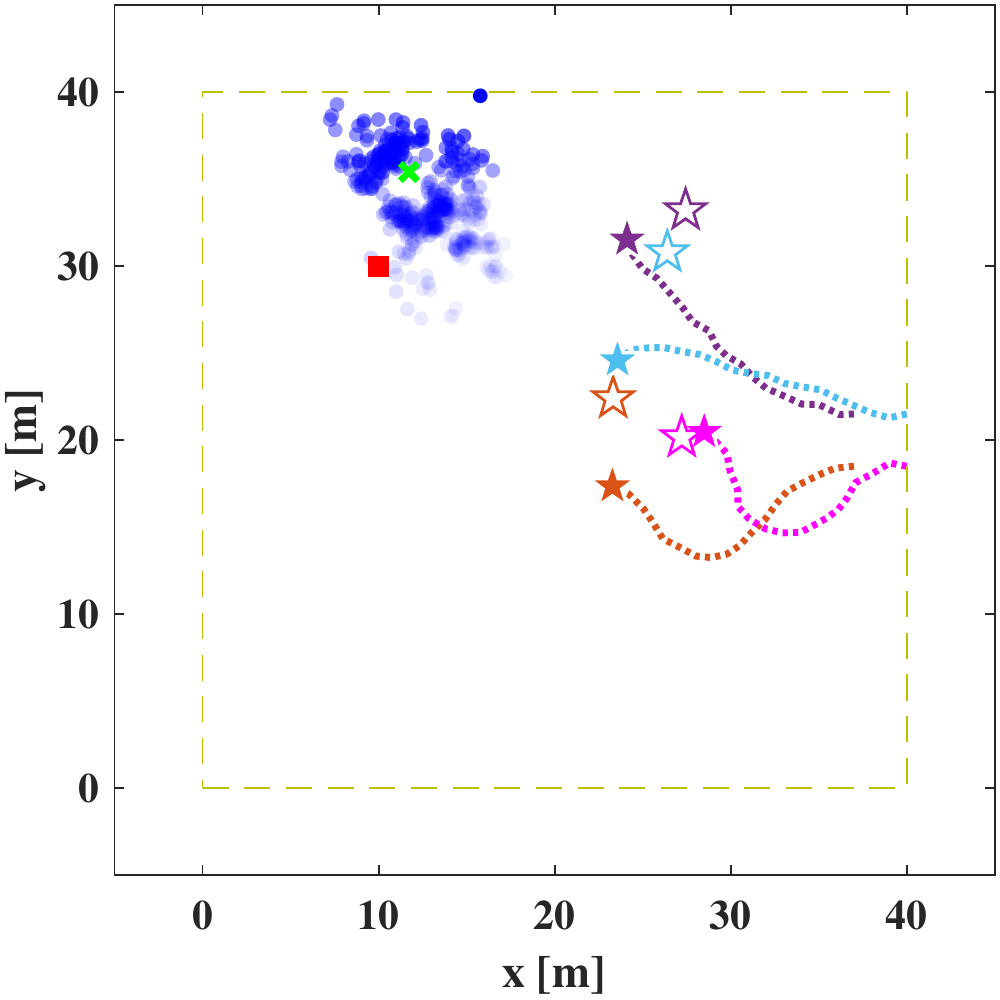}
         \caption{$t=17s$}
     \end{subfigure}
     \begin{subfigure}[b]{.24\textwidth}
         \centering
         \includegraphics[width=\textwidth]{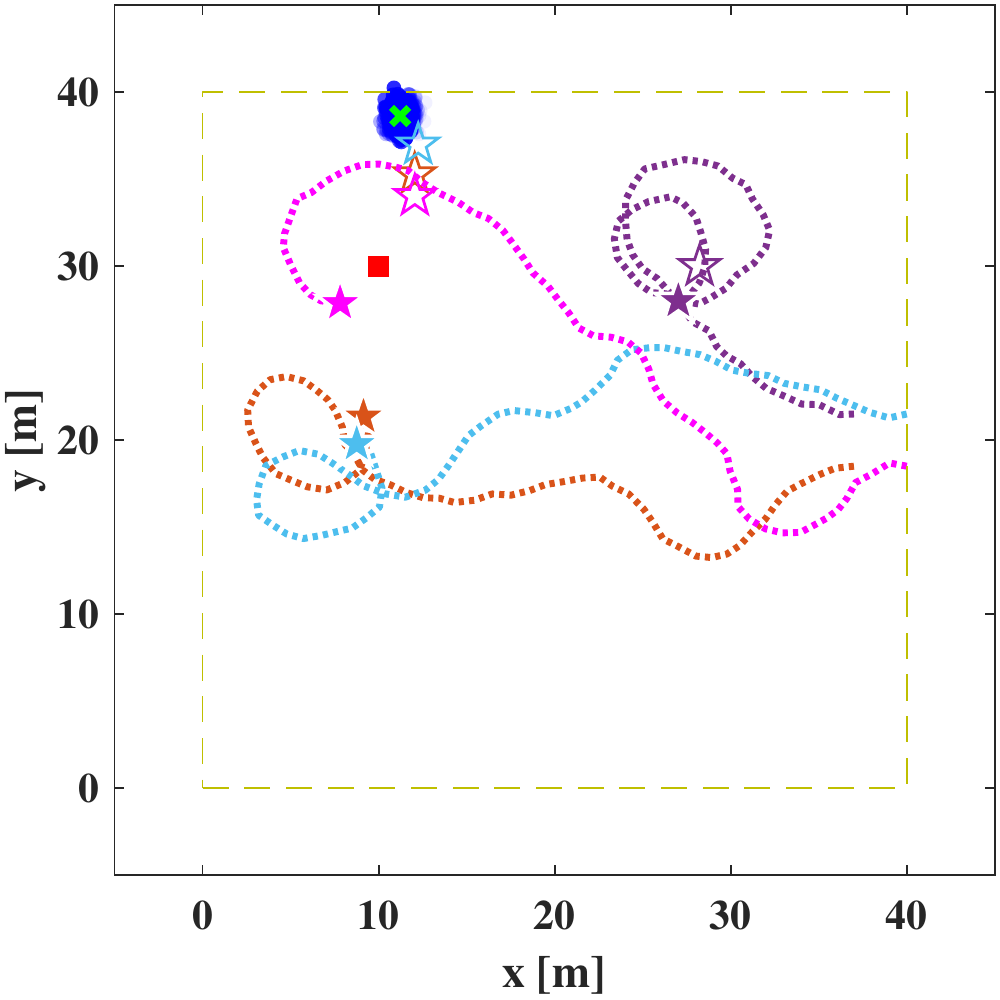}
         \caption{$t=55s$}
     \end{subfigure}\\
     \begin{subfigure}[b]{.24\textwidth}
         \centering
         \includegraphics[width=\textwidth]{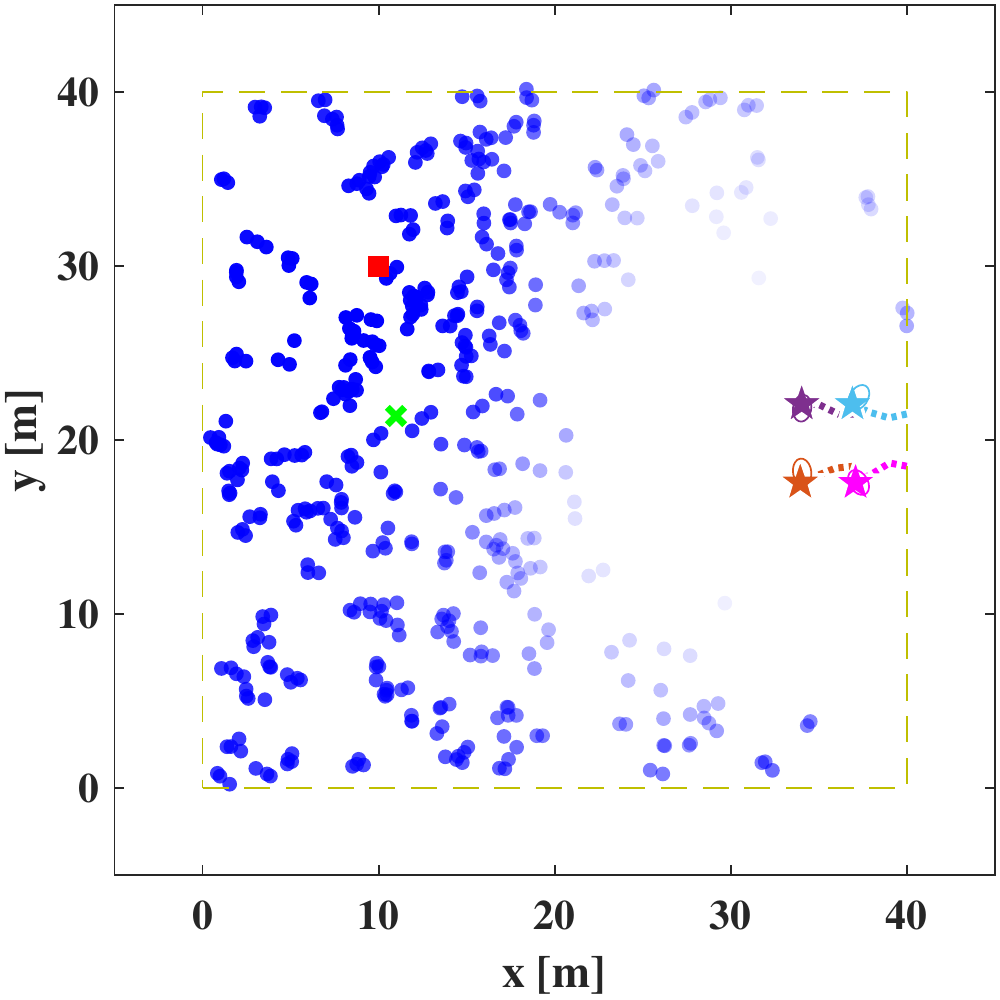}
         \caption{$t=3s$}
     \end{subfigure}
     \begin{subfigure}[b]{.24\textwidth}
         \centering
         \includegraphics[width=\textwidth]{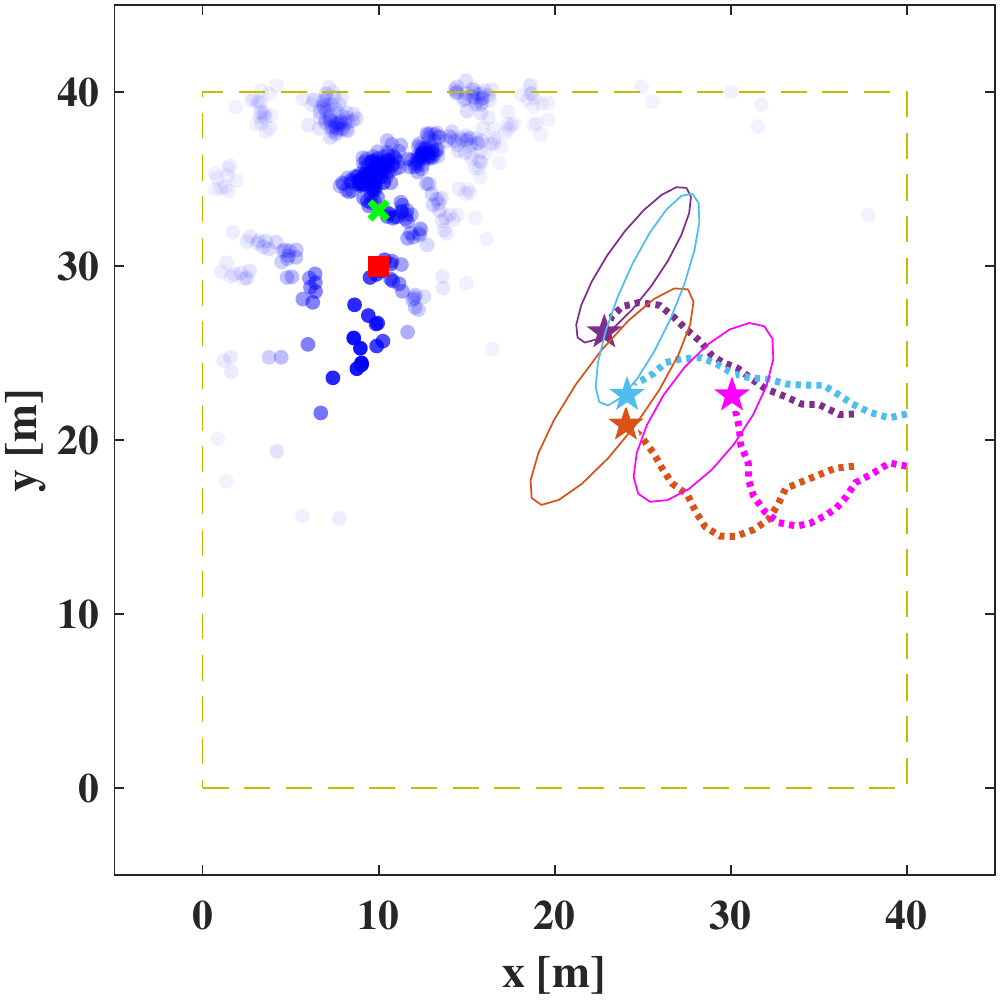}
         \caption{$t=17s$}
     \end{subfigure}
     \begin{subfigure}[b]{.24\textwidth}
         \centering
         \includegraphics[width=\textwidth]{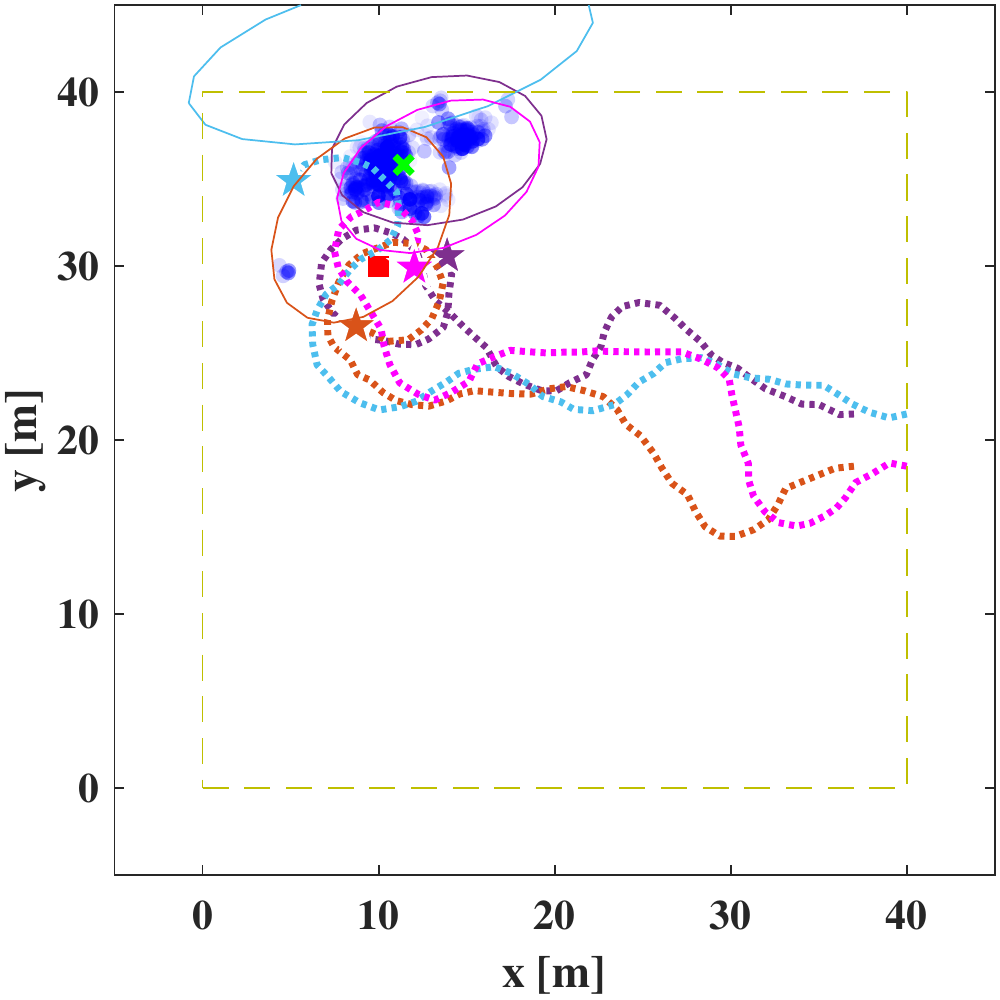}
         \caption{$t=55s$}
     \end{subfigure}
    \caption{Snapshots and resulting trajectories from (a)-(c) PF-only method, and (d)-(f) the proposed algorithm with noise covariance 6$P_0$}
    \label{fig:6P}
% \end{minipage}
\end{figure}

%\section*{Acknowledgments}
%An Acknowledgments section, if used, \textbf{immediately precedes} the References. Sponsorship information and funding data are included here. The preferred spelling of the word ``acknowledgment'' in American English is without the ``e'' after the ``g.'' Avoid expressions such as ``One of us (S.B.A.) would like to thank\ldots'' Instead, write ``F.~A.~Author thanks\ldots'' Sponsor and financial support acknowledgments are also to be listed in the ``acknowledgments'' section.

\clearpage

\bibliography{sample}

\begin{thebibliography}{22}
\newcommand{\enquote}[1]{``#1''}
\providecommand{\natexlab}[1]{#1}
\providecommand{\url}[1]{\texttt{#1}}
\providecommand{\urlprefix}{URL }
\expandafter\ifx\csname urlstyle\endcsname\relax
  \providecommand{\doi}[1]{doi:\discretionary{}{}{}#1}\else
  \providecommand{\doi}{doi:\discretionary{}{}{}\begingroup
  \urlstyle{rm}\Url}\fi

\bibitem[{Dunbabin and Marques(2012)}]{dunbabin2012robots}
Dunbabin, M., and Marques, L., \enquote{Robots for environmental monitoring:
  Significant advancements and applications,} \emph{IEEE Robotics \& Automation
  Magazine}, Vol.~19, No.~1, 2012, pp. 24--39.

\bibitem[{Choi and How(2010)}]{choi2010continuous}
Choi, H.-L., and How, J.~P., \enquote{Continuous trajectory planning of mobile
  sensors for informative forecasting,} \emph{Automatica}, Vol.~46, No.~8,
  2010, pp. 1266--1275.

\bibitem[{Hoffmann and Tomlin(2009)}]{hoffmann2009mobile}
Hoffmann, G.~M., and Tomlin, C.~J., \enquote{Mobile sensor network control
  using mutual information methods and particle filters,} \emph{IEEE
  Transactions on Automatic Control}, Vol.~55, No.~1, 2009, pp. 32--47.

\bibitem[{Shen et~al.(2010)Shen, Chen, Blasch, Pham, Douville, Yang, and
  Kadar}]{shen2010game}
Shen, D., Chen, G., Blasch, E., Pham, K., Douville, P., Yang, C., and Kadar,
  I., \enquote{Game theoretic sensor management for target tracking,}
  \emph{Signal Processing, Sensor Fusion, and Target Recognition XIX}, Vol.
  7697, International Society for Optics and Photonics, 2010, p. 76970C.

\bibitem[{Choi and Lee(2015)}]{choi2015potential}
Choi, H.-L., and Lee, S.-J., \enquote{A potential-game approach for
  information-maximizing cooperative planning of sensor networks,} \emph{IEEE
  Transactions on Control Systems Technology}, Vol.~23, No.~6, 2015, pp.
  2326--2335.

\bibitem[{Lee et~al.(2018)Lee, Park, and Choi}]{lee2018potential}
Lee, S.-J., Park, S.-S., and Choi, H.-L., \enquote{Potential Game-Based
  Non-Myopic Sensor Network Planning for Multi-Target Tracking,} \emph{IEEE
  Access}, Vol.~6, 2018, pp. 79245--79257.

\bibitem[{Kolmogorov(1956)}]{kolmogorov1956shannon}
Kolmogorov, A., \enquote{On the Shannon theory of information transmission in
  the case of continuous signals,} \emph{IRE Transactions on Information
  Theory}, Vol.~2, No.~4, 1956, pp. 102--108.

\bibitem[{Misra and Enge(2006)}]{misra2006global}
Misra, P., and Enge, P., \enquote{Global Positioning System: signals,
  measurements and performance second edition,} \emph{Massachusetts:
  Ganga-Jamuna Press}, 2006.

\bibitem[{Balamurugan et~al.(2016)Balamurugan, Valarmathi, and
  Naidu}]{balamurugan2016survey}
Balamurugan, G., Valarmathi, J., and Naidu, V., \enquote{Survey on UAV
  navigation in GPS denied environments,} \emph{2016 International Conference
  on Signal Processing, Communication, Power and Embedded System (SCOPES)},
  IEEE, 2016, pp. 198--204.

\bibitem[{Hardy et~al.(2016)Hardy, Strader, Gross, Gu, Keck, Douglas, and
  Taylor}]{hardy2016unmanned}
Hardy, J., Strader, J., Gross, J.~N., Gu, Y., Keck, M., Douglas, J., and
  Taylor, C.~N., \enquote{Unmanned aerial vehicle relative navigation in GPS
  denied environments,} \emph{2016 IEEE/ION Position, Location and Navigation
  Symposium (PLANS)}, IEEE, 2016, pp. 344--352.

\bibitem[{Hemann et~al.(2016)Hemann, Singh, and Kaess}]{hemann2016long}
Hemann, G., Singh, S., and Kaess, M., \enquote{Long-range GPS-denied aerial
  inertial navigation with LIDAR localization,} \emph{2016 IEEE/RSJ
  International Conference on Intelligent Robots and Systems (IROS)}, IEEE,
  2016, pp. 1659--1666.

\bibitem[{Achtelik et~al.(2009)Achtelik, Bachrach, He, Prentice, and
  Roy}]{achtelik2009autonomous}
Achtelik, M., Bachrach, A., He, R., Prentice, S., and Roy, N.,
  \enquote{Autonomous navigation and exploration of a quadrotor helicopter in
  GPS-denied indoor environments,} \emph{First Symposium on Indoor Flight},
  Citeseer, 2009.

\bibitem[{Kassas and Humphreys(2013)}]{kassas2013observability}
Kassas, Z.~M., and Humphreys, T.~E., \enquote{Observability analysis of
  collaborative opportunistic navigation with pseudorange measurements,}
  \emph{IEEE Transactions on Intelligent Transportation Systems}, Vol.~15,
  No.~1, 2013, pp. 260--273.

\bibitem[{He et~al.(2008)He, Prentice, and Roy}]{he2008planning}
He, R., Prentice, S., and Roy, N., \enquote{Planning in information space for a
  quadrotor helicopter in a GPS-denied environment,} \emph{2008 IEEE
  International Conference on Robotics and Automation}, IEEE, 2008, pp.
  1814--1820.

\bibitem[{Montemerlo et~al.(2002)Montemerlo, Thrun, Koller, Wegbreit
  et~al.}]{montemerlo2002fastslam}
Montemerlo, M., Thrun, S., Koller, D., Wegbreit, B., et~al., \enquote{FastSLAM:
  A factored solution to the simultaneous localization and mapping problem,}
  \emph{Aaai/iaai}, Vol. 593598, 2002.

\bibitem[{Chen et~al.(2003)}]{chen2003bayesian}
Chen, Z., et~al., \enquote{Bayesian filtering: From Kalman filters to particle
  filters, and beyond,} \emph{Statistics}, Vol. 182, No.~1, 2003, pp. 1--69.

\bibitem[{Murphy(2000)}]{murphy2000bayesian}
Murphy, K.~P., \enquote{Bayesian map learning in dynamic environments,}
  \emph{Advances in Neural Information Processing Systems}, 2000, pp.
  1015--1021.

\bibitem[{Thrun et~al.(2005)Thrun, Burgard, and Fox}]{thrun2005probabilistic}
Thrun, S., Burgard, W., and Fox, D., \emph{Probabilistic robotics}, MIT press,
  2005.

\bibitem[{Skolnik(1970)}]{skolnik1970radar}
Skolnik, M.~I., \enquote{Radar handbook,} 1970.

\bibitem[{Williams et~al.(2007)Williams, Fisher, and
  Willsky}]{williams2007approximate}
Williams, J.~L., Fisher, J.~W., and Willsky, A.~S., \enquote{Approximate
  dynamic programming for communication-constrained sensor network management,}
  \emph{IEEE Transactions on signal Processing}, Vol.~55, No.~8, 2007, pp.
  4300--4311.

\bibitem[{Lee and Choi(2014)}]{lee2014efficient}
Lee, S.-J., and Choi, H.-L., \enquote{An efficient particle filter-based
  potential game method for distributed sensor network management,}
  \emph{SENSORS, 2014 IEEE}, IEEE, 2014, pp. 1256--1259.

\bibitem[{Owen et~al.(2015)Owen, Beard, and McLain}]{owen2015implementing}
Owen, M., Beard, R.~W., and McLain, T.~W., \enquote{Implementing dubins
  airplane paths on fixed-wing uavs,} \emph{Handbook of Unmanned Aerial
  Vehicles}, 2015, pp. 1677--1701.

\end{thebibliography}

\end{document}